%% file: main.tex
\documentclass[final,5p,times,twocolumn, number, sort&compress]{elsarticle}

\usepackage{amssymb}
\usepackage{lipsum}
\usepackage{graphicx}% Include figure files
\usepackage{bm}% bold math
\usepackage{subcaption}
\usepackage{mathtools}
\usepackage{comment}
\usepackage[style=plain]{floatrow}
\usepackage{xspace}
\usepackage[dvipsnames]{xcolor}
\usepackage[colorlinks=true]{hyperref}

\allowdisplaybreaks
\journal{Physics Letters B}
\newcommand{\code}[1]{\texttt{#1}}
\newcommand{\dd}{\text{d}}
\newcommand{\ycut}{y_\text{cut}}

\begin{document}

\begin{frontmatter}

\title{Jet production at electron-positron colliders \\
at next-to-next-to-next-to-leading order in QCD}

\author[first]{Xuan Chen}
\affiliation[first]{organization={School of Physics, Shandong University},%Department and Organization
            %addressline={}, 
            city={Shandong},
            postcode={250100}, 
            state={Jinan},
            country={China}}
            
\author[second]{Petr Jakub\v{c}\'{i}k}
\affiliation[second]{organization={Physik-Institut, Universit\"at Z\"urich},%Department and Organization
            addressline={Winterthurerstrasse 190}, 
            city={Z\"urich},
            postcode={CH-8057}, 
            %state={},
            country={Switzerland}}
            
\author[third]{Matteo Marcoli}
\affiliation[third]{organization={Institute for Particle Physics Phenomenology, Department of Physics, Durham University},%Department and Organization
            %addressline={}, 
            city={Durham},
            postcode={DH1 3LE}, 
            %state={},
            country={UK}}
            
\author[fourth]{Giovanni Stagnitto}
\affiliation[fourth]{organization={Universit\`{a} degli Studi di Milano-Bicocca \& INFN},%Department and Organization
            addressline={Piazza della Scienza 3}, 
            city={Milano},
            postcode={20126}, 
            %state={},
            country={Italy}}

%\author{Xuan Chen}
%\affiliation{School of Physics, Shandong University, Jinan, Shandong 250100, China}
%\author{Petr Jakub\v{c}\'{i}k}
%\affiliation{Physik-Institut, Universit\"at Z\"urich,Winterthurerstrasse 190, CH-8057 Z\"urich, Switzerland}
%\author{Matteo Marcoli}
%\affiliation{Institute for Particle Physics Phenomenology, Department of Physics, Durham University, Durham, DH1 3LE, UK}
%\author{Giovanni Stagnitto}
%\affiliation{Universit\`{a} degli Studi di Milano-Bicocca \& INFN, Piazza della Scienza 3, Milano 20126, Italy}

\begin{abstract}
We present the first application of antenna subtraction at next-to-next-to-next-to-leading order (N$^3$LO) in QCD by computing fully differential predictions for two-jet production at electron-positron colliders.
We illustrate the structure of the infrared counterterms and provide results for the N$^3$LO correction to the two-jet production rate and to the leading-jet energy. % in two-jet events.
Our work constitutes the first direct calculation of jet production at electron-positron colliders at N$^3$LO and represents the first step in tackling arbitrary processes with jets at this perturbative order.
\end{abstract}

%\begin{keyword}
%% keywords here, in the form: keyword \sep keyword, up to a maximum of 6 keywords
%keyword 1 \sep keyword 2 \sep keyword 3 \sep keyword 4



%\end{keyword}

\end{frontmatter}

\input{Sections/introduction}

\input{Sections/method}

\input{Sections/results}

\input{Sections/conclusion}

\section*{Acknowledgements}
We are grateful to Thomas Gehrmann, Nigel Glover and Aude Gehrmann-de Ridder for support and encouragement to pursue this work.
We thank Elliot Fox for comments on the manuscript.
XC is supported by the National Science Foundation of China (NSFC) with grant No.12475085 and No.12321005.
PJ is supported
by the European Research Council (ERC) under the European Union’s research and innovation programme grant agreement 101019620 (ERC Advanced Grant TOPUP).
MM is supported by a Royal Society Newton International Fellowship (NIF/R1/232539).
We acknowledge financial support, supercomputing resources and support from ICSC – Centro Nazionale di Ricerca in High Performance Computing, Big Data and Quantum Computing – and hosting entity, funded by European Union – NextGenerationEU.
We are grateful to each other's institutions and the Pauli Center for Theoretical Studies for hospitality.

\bibliographystyle{elsarticle-harv} 
\bibliography{main}

\end{document}

%% file: Sections/introduction.tex
\section{Introduction}
\vspace{-6pt}
Experiments at present and future colliders plan to deliver an unprecedented amount of data and significantly increase our sensitivity to new physics if equally precise Standard Model predictions are available.
In high-energy collisions, the effects of QCD are dominant and their evaluation at higher perturbative orders in the strong coupling $\alpha_s$ requires techniques for the identification and removal of infrared (IR) divergences from the constituent scattering amplitudes, known as \textit{infrared subtraction schemes}.
%Figure here to force it on top of page 2

\input{Figures/figure1.tex}

General approaches have been developed for next-to-leading order (NLO) calculations~\cite{Catani:1996vz,Frixione:1995ms}, followed by several strategies at next-to-next-to-leading order (NNLO)~\cite{Gehrmann-DeRidder:2005btv,Boughezal:2011jf,Currie:2013vh,DelDuca:2016ily,Catani:2007vq,Czakon:2010td,Czakon:2014oma,Gaunt:2015pea,Cacciari:2015jma,Caola:2017dug,Magnea:2018hab,Herzog:2018ily,TorresBobadilla:2020ekr,Bertolotti:2022aih} and ongoing efforts to include higher-multiplicity final states and to generalize the current formalisms~\cite{Gehrmann:2023dxm,Fox:2024bfp,Bonino:2024adk,Devoto:2023rpv,Devoto:2025kin}.
NNLO-accurate predictions are nowadays available for several Standard Model processes, see~\cite{Huss:2022ful} for a review. 
N$^{3}$LO accuracy has recently been reached for color-singlet production and decay processes for both inclusive~\cite{Anastasiou:2015vya,Dreyer:2016oyx,Mistlberger:2018etf,Dreyer:2018qbw,Chen:2019lzz,Chen:2019fhs,Baglio:2022wzu,Duhr:2021vwj, Duhr:2020seh,Duhr:2020sdp} and differential~\cite{Chen:2022lwc,Campbell:2023lcy,Chen:2021vtu,Chen:2022cgv,Neumann:2022lft,Cieri:2018oms,Chen:2021isd,Billis:2021ecs,Mondini:2019gid,Fox:2025cuz,He:2025hin} observables,
relying either on $q_T$-slicing~\cite{Catani:2007vq} or the Projection-to-Born~\cite{Cacciari:2015jma} method.
Neither of these approaches is applicable to arbitrary processes and observables, especially in the presence of final-state jets, making the development of a general N$^3$LO subtraction scheme highly desirable.

The \textit{antenna subtraction} method, formulated at NLO in~\cite{Kosower:1997zr,Campbell:1998nn} and extended to NNLO in~\cite{Gehrmann-DeRidder:2005btv,Currie:2013vh} is one of the most flexible local subtraction schemes with the ability to handle jets at both lepton and hadron colliders. 
Its~fundamental ingredients are \textit{antenna functions}~\cite{Gehrmann-DeRidder:2004ttg,Gehrmann-DeRidder:2005svg,Gehrmann-DeRidder:2005alt}, which encode the singular behavior of real-emission matrix elements and can be analytically integrated over the phase space of the unresolved radiation in order to cancel the explicit IR singularities of virtual corrections. 

In this Letter, we present the first application of antenna subtraction to fully differential predictions for physical cross sections at N$^3$LO.
We focus on the process $e^+e^-\to jj + X$, which was computed at NNLO with the sector decomposition~\cite{Anastasiou:2004qd} and antenna subtraction~\cite{Weinzierl:2006ij} methods.
In~\cite{Gehrmann-DeRidder:2008qsl}, the N$^3$LO correction to the two-jet production rate was evaluated differentially in the Durham jet reconstruction algorithm parameter $\ycut$~\cite{Catani:1991hj,Brown:1990nm,Brown:1991hx,Stirling:1991ds,Bethke:1991wk} indirectly, i.e. by combining an NNLO calculation for three-jet production with the analytic result for the N$^3$LO total inclusive cross section for $e^+e^-\to$ hadrons~\cite{Chetyrkin:1996ela}. 
In this Letter, we evaluate the two-jet rate with a direct method that can in principle account for any experimental cuts.
In addition, we determine the distribution of the leading-jet energy in two-jet events, previously unknown at $\mathcal{O}(\alpha_s^3)$. 

This work represents the first N$^3$LO differential calculation obtained with a subtraction scheme that is not restricted to specific classes of processes and does not rely on the prior knowledge of inclusive quantities. 
Our calculation is therefore a crucial step in tackling more complicated processes with jets at high perturbative order~\cite{Caola:2022ayt} relevant to the physics goals of the upcoming High-Luminosity phase of the LHC~\cite{ZurbanoFernandez:2020cco}.
Moreover, results at this level of precision are relevant for future QCD precision studies~\cite{Verbytskyi:2025sod,2921876} at the next generation of electron-positron colliders, such as the FCC-ee~\cite{FCC:2018byv,FCC:2018evy,FCC:2025lpp} and CEPC~\cite{An:2018dwb,CEPCPhysicsStudyGroup:2022uwl,Ai:2024nmn}.

%% file: Figures/figure1.tex
\begin{figure*}[t]
\centering
\begin{subfigure}{.21\textwidth}
  \centering
  \includegraphics[width=\textwidth]{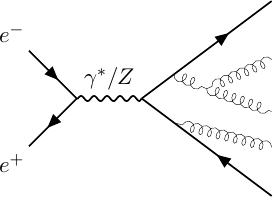}
  \caption{triple-real}
  \label{fig11:RRR}
\end{subfigure}
\hspace{16pt}
\begin{subfigure}{.21\textwidth}
  \centering
  \includegraphics[width=\textwidth]{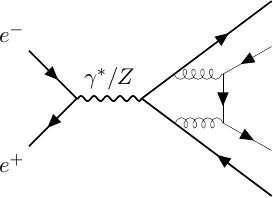}
  \caption{double-real-virtual}
  \label{fig1b:RRV}
\end{subfigure}
\hspace{16pt}
\begin{subfigure}{.21\textwidth}
  \centering
  \includegraphics[width=\textwidth]{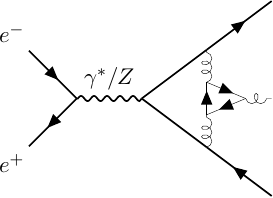}
  \caption{real-double-virtual}
  \label{fig1c:RVV}
\end{subfigure}
\hspace{16pt}
\begin{subfigure}{.21\textwidth}
  \centering
  \includegraphics[width=\textwidth]{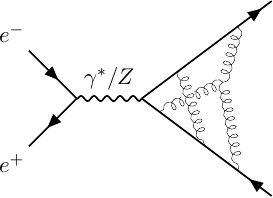}
  \caption{triple-virtual}
  \label{fig1d:VVV}
\end{subfigure}
\caption{Representative Feynman diagrams (before interference with an appropriate tree-level diagram) for each type of radiative correction to the process $e^+e^-\to2\,$jets at $\mathcal{O}(\alpha_s^3)$. From left to right: $\dd\sigma^{(3)}_5$, $\dd\sigma^{(3)}_4$, $\dd\sigma^{(3)}_3$, and $\dd\sigma^{(3)}_2$.}
\label{fig1:diagrams}
\end{figure*}

%% file: Sections/method.tex
\section{Method}
In the following, we outline the extension of antenna subtraction for final-state radiation to N$^3$LO.
The $\mathcal{O}(\alpha_s^3)$ correction to two-jet production at electron-positron colliders receives contributions from final states with up to five partons.
We denote the contribution to the N$^3$LO coefficient from a configuration with $n$ final-state partons by $\dd\sigma^{(3)}_n$.
The \textit{triple-real} contribution $\dd\sigma^{(3)}_5$ is given by three real emissions on top of the Born-level process, the \textit{double-real-virtual} contribution $\dd\sigma^{(3)}_4$ by two real emissions at one loop, the \textit{real-double-virtual} contribution $\dd\sigma^{(3)}_3$ by one real emission at two loops and finally the \textit{triple-virtual} contribution $\dd\sigma^{(3)}_2$ by the three-loop correction.
Examples of Feynman diagrams contributing to different types of N$^3$LO corrections are shown in Figure~\ref{fig1:diagrams}.

The individual contributions contain explicit IR singularities due to loop corrections ($\dd\sigma^{(3)}_2$), implicit singularities arising when the external momenta approach soft and collinear limits ($\dd\sigma^{(3)}_5$), or a mixture of both ($\dd\sigma^{(3)}_3$ and $\dd\sigma^{(3)}_4$).
These divergences cancel when virtual and real corrections are combined in physical predictions~\cite{Kinoshita:1962ur,Lee:1964is}, but each individual term is numerically integrable over the associated phase space only after introducing dedicated subtraction terms.
We denote by $\dd\sigma^{(3)}_{n,\text{finite}}$ the individual contributions after subtraction, free from any type of IR singularity.
The definition of $\dd\sigma^{(k)}_{n,\text{finite}}$ up to NNLO in the context of antenna subtraction was described in~\cite{Gehrmann-DeRidder:2005btv,Currie:2013vh}.
At N$^3$LO, we write schematically
\begin{alignat}{6}
    &\label{eq:layer1}\dd\sigma_{5,\text{finite}}^{(3)} &&= \dd\sigma_{5}^{(3)} &&- \dd\sigma^{(3)}_{5,S_{1}} &&\hspace{-2pt}- \dd\sigma^{(3)}_{5,S_{2}} &&\hspace{-4pt}- \dd\sigma^{(3)}_{5,S_{3}}\,, \\
    &\label{eq:layer2}\dd\sigma_{4,\text{finite}}^{(3)} &&= \dd\sigma_{4}^{(3)} &&- \dd\sigma^{(3)}_{4,V_{1}} &&\hspace{-2pt}- \dd\sigma^{(3)}_{4,V_{1}S_{1}} &&\hspace{-4pt}- \dd\sigma^{(3)}_{4,V_{1}S_{2}}\,, \\
    &\label{eq:layer3}\dd\sigma_{3,\text{finite}}^{(3)} &&= \dd\sigma_{3}^{(3)} &&- \dd\sigma^{(3)}_{3,V_{2}} &&\hspace{-2pt}- \dd\sigma^{(3)}_{3,V_{2}S_{1}}\,,
    \\
    &\label{eq:layer4}\dd\sigma_{2,\text{finite}}^{(3)} &&= \dd\sigma_{2}^{(3)} &&- \dd\sigma^{(3)}_{2,V_{3}}\,,
\end{alignat}
where $\dd\sigma^{(3)}_{n,V_{\ell}S_{m}}$ represents a counterterm targeting the divergent behavior of $m$ unresolved partons as well as the explicit IR poles in a $\ell$-loop matrix element.

The following relations between the subtraction terms ensure the cancellation of IR singularities across different phase-space multiplicities:
\begin{eqnarray}
    \label{eq:closure1}\dd\sigma^{(3)}_{4,V_{1}} &=& -\hspace{-2pt}\int_1\dd\sigma^{(3)}_{5,S_{1}}\,, \\
    \label{eq:closure2}\dd\sigma^{(3)}_{3,V_{2}} &=& -\hspace{-2pt}\int_1\dd\sigma^{(3)}_{4,V_{1}S_{1}}\hspace{-2pt}-\hspace{-2pt}\int_2\dd\sigma^{(3)}_{5,S_{2}}\,, \\
    \label{eq:closure3}\hspace{-12pt}\dd\sigma^{(3)}_{2,V_{3}} &=& -\hspace{-2pt}\int_1\dd\sigma^{(3)}_{3,V_{2}S_{1}}\hspace{-2pt}-\hspace{-2pt}\int_2\dd\sigma^{(3)}_{4,V_{1}S_{2}}\hspace{-2pt}-\hspace{-2pt}\int_3\dd\sigma^{(3)}_{5,S_{3}}\,,
\end{eqnarray}
where $\int_m$ indicates inclusive integration over the phase space of $m$ unresolved emissions. As a consequence of the relations~\eqref{eq:closure1}--\eqref{eq:closure3}, the net contribution of the subtraction terms vanishes in the sum of~\eqref{eq:layer1}--\eqref{eq:layer4}, restoring the correct physical result
\begin{equation}
    \sum_{n=2}^5\int_n\dd\sigma^{(3)}_{n,\text{finite}}=\sum_{n=2}^5\int_n\dd\sigma^{(3)}_n\,,
\end{equation}
with the crucial difference that each term in the sum on the left-hand side is individually well-defined and computable with numerical techniques.

The subtraction terms $\dd\sigma^{(3)}_{5,S_{1}}$, $\dd\sigma^{(3)}_{5,S_{2}}$, $\dd\sigma^{(3)}_{4,V_{1}}$, $\dd\sigma^{(3)}_{4,V_{1}S_{1}}$ and $\dd\sigma^{(3)}_{3,V_{2}}$ are analogous to those required for an NNLO calculation of three-jet production in electron-positron annihilation~\cite{Gehrmann-DeRidder:2007foh,Weinzierl:2009nz}.
The subtraction terms $\dd\sigma^{(3)}_{5,S_{3}}$, $\dd\sigma^{(3)}_{4,V_{1}S_{2}}$, $\dd\sigma^{(3)}_{3,V_{2}S_{1}}$ and $\dd\sigma^{(3)}_{2,V_{3}}$ specifically target N$^3$LO infrared singularities and have been constructed for the first time for the present calculation.

The subtraction terms are built from antenna functions with the following notation: $X_n^\ell$ is a $n$-parton $\ell$-loop antenna function intended to subtract the singular behavior of $n-2$ unresolved partons between a pair of hard radiators at $\ell$ loops.
$X$ is a placeholder for a particular partonic configuration~\cite{Gehrmann-DeRidder:2005btv}.
The quark-antiquark antenna functions relevant for our process are obtained from color-ordered matrix elements for the decay of a photon to $n$ partons~\cite{Gehrmann-DeRidder:2004ttg, Jakubcik:2022zdi}, denoted by $M_{n,\gamma^*}^\ell$.
The integrated counterpart of an antenna function, $\mathcal{X}_n^\ell$, is obtained by analytic integration over the inclusive phase space of the $n-2$ unresolved emissions.
All the N$^3$LO final-state antenna functions were recently integrated in~\cite{Jakubcik:2022zdi,Chen:2023fbaHiggs,Chen:2023egxNeutralino}.

The infrared limits appearing at N$^3$LO cannot in general be addressed by a combination of NLO and NNLO ingredients, and three new types antenna functions must be defined: five-parton tree-level
\begin{align}
\label{eq:X50}
    X_5^0&=N_5\,\dfrac{M_{5,\gamma^*}^0}{M_{2,\gamma^*}^0}\,,
\end{align}
four-parton one-loop
\begin{align}
      \label{eq:X41}
    X_4^1&=N_4\,\dfrac{M_{4,\gamma^*}^1}{M_{2,\gamma^*}^0}-X_4^0\dfrac{M_{2,\gamma^*}^1}{M_{2,\gamma^*}^0}\,,
\end{align}
and three-parton two-loop
\begin{align}
    \label{eq:X32}
    X_3^2&=N_3\dfrac{M_{3,\gamma^*}^2}{M_{2,\gamma^*}^0}-X_3^1\dfrac{M_{2,\gamma^*}^1}{M_{2,\gamma^*}^0}-X_3^0\dfrac{M_{2,\gamma^*}^2}{M_{2,\gamma^*}^0}\,,
\end{align}
where $N_i$ indicates a numerical symmetry factor and the two-body decay matrix element $M_{2,\gamma^*}^0$ is a normalization factor.
In the case of the loop antenna functions $X_4^1$ and $X_3^2$, we remove
the singular behavior that can be captured by antenna functions of lower loop order, naturally extending the definition of three-parton one-loop antenna functions at NNLO~\cite{Gehrmann-DeRidder:2005btv}.

The $X_5^0$ antenna functions are used in the subtraction term $\dd\sigma^{(3)}_{5,S_{3}}$ to remove triple-real unresolved configurations.
The singular behavior of the double-real-virtual correction is addressed with the functions $X_4^1$ within the term $\dd\sigma^{(3)}_{4,V_{1}S_{2}}$, while $X_3^2$ are used in $\dd\sigma^{(3)}_{3,V_{2}S_{1}}$ to subtract the soft and collinear divergences of the real-double-virtual correction.
Finally, the term $\dd\sigma^{(3)}_{2,V_{3}}$ collects the integrated counterparts of all the N$^3$LO antenna functions ($\mathcal{X}_5^0$, $\mathcal{X}_4^1$, $\mathcal{X}_3^2$) and removes the explicit IR singularities of the triple-virtual correction.
Throughout, we work with spin-averaged antenna functions and we fully remove azimuthal correlations in collinear limits using the angular averaging routines described in~\cite{NigelGlover:2010kwr}, extended to handle triple-unresolved configurations.

After the removal of IR poles, the construction of $\dd\sigma^{(3)}_{3,V_{2}S_{1}}$ and $\dd\sigma^{(3)}_{4,V_{1}S_{2}}$ closely follows the logic of single-real and double-real infrared subtraction, already considered up to NNLO.
A subtlety arises when a one-loop reduced matrix element and a one-loop unresolved factor appear in a product.
To eliminate the need for higher orders in the dimensional regulator $\epsilon$ in one-loop matrix elements, we consistently work with IR finite remainders.

Let us elaborate on a crucial aspect of a N$^3$LO local subtraction scheme---the removal of triple-unresolved limits.
We start by recalling the structure of single-unresolved and double-unresolved subtraction terms:
\begin{alignat}{3}
    \label{dsigmaS1} \dd\sigma_{5,S_{1}}^{(3)} &= &&X_3^0 \otimes M_{4}^0 J_2^{(4)}, &&\tag{12}\\
    \dd\sigma_{5,S_{2}}^{(3)} &= \Big[&&\phantom{\otimes\,}\sum_{cc} (X_4^0 - X_3^0 \otimes X_3^0 ) \tag{13a}\label{RR:term1} \\[-3pt]
    & &&+\sum_{acc}(X_3^0 \otimes X_3^0 + \mathcal{S} \otimes X_3^0)\Big]  \tag{13b}\label{RR:term2}\\[-3pt]
    & &&\otimes M_{3}^0 J_2^{(3)}, \nonumber
\end{alignat}
where $M_n^0$ represents a generic tree-level color-ordered matrix element for $e^+ e^-\to n$ partons and the shorthand symbol $\otimes$ encodes a momentum mapping from a higher-multiplicity to a lower-multiplicity phase space, in order to absorb unresolved emissions. For each term, a sum over external partons and color correlations is understood.
The jet function $J_{n_j}^{(n_p)}$ reconstructs $n_j$ jets from $n_p$ final-state partons.
The sums in~\eqref{RR:term1} and~\eqref{RR:term2} run over configurations where the two unresolved emissions are respectively \textit{color-connected} ($cc$) and \textit{almost color-connected} ($acc$)~\cite{Gehrmann-DeRidder:2005btv,Gehrmann-DeRidder:2007foh}.

The singular behavior of color-connected emissions can only be captured by dedicated $X_4^0$ antenna functions.
To prevent double-counting of IR limits already addressed by~\eqref{dsigmaS1}, we introduce the term $X_3^0\otimes X_3^0$ in~\eqref{RR:term1}, which subtracts the single-unresolved singularities from the $X_4^0$.
In the second term of~\eqref{RR:term2}, $\mathcal{S}$ represents a combination of eikonal factors required to remove soft singularities at large angle~\cite{Weinzierl:2008iv,Gehrmann-DeRidder:2007foh}.
Finally, the $X_3^0\otimes X_3^0$ and $\mathcal{S}\otimes X_3^0$ structures present in $\dd\sigma_{5,S_{2}}^{(3)}$ cancel against $\dd\sigma_{5,S_{1}}^{(3)}$ in~\eqref{dsigmaS1} when the reduced matrix element $M_{4}^0$ develops an additional singularity on top of that already captured by the $X_3^0$ antenna function.
Together, $\dd\sigma_{5,S_{1}}^{(3)}+\dd\sigma_{5,S_{2}}^{(3)}$ subtract all the single- and double-unresolved behavior of the five-parton tree-level matrix element fully differentially across the phase space.

The triple-real subtraction terms have an analogous structure:
\begin{align}
    \dd\sigma_{5,S_{3}}^{(3)} = \Big[&\phantom{-}X_5^0 \tag{14a} \label{RRR:term1}\\[-3pt]
    &- X_3^0 \otimes X_4^0 \tag{14b} \label{RRR:term2}\\[-3pt]
    &- \sum_{cc} (X_4^0-X_3^0\otimes X_3^0) \otimes X_3^0 \tag{14c}\label{RRR:term3}\\[-3pt]
    &-\sum_{acc} (X_3^0\otimes X_3^0+\mathcal{S} \otimes X_3^0) \otimes X_3^0\Big] \tag{14d}\label{RRR:term4}\\[-3pt]
    &\phantom{-}\otimes M_{2}^0 J_2^{(2)} \nonumber
\end{align}
Here the $X_5^0$ antenna functions address color-connected  triple-unresolved emissions.
In particular, they contain the singular behavior in triple-soft~\cite{Catani:2019nqv,DelDuca:2022noh,Catani:2022hkb} and quadruple-collinear~\cite{DelDuca:2019ggv,DelDuca:2020vst} limits.
A $5\to2$ antenna momentum mapping~\cite{Kosower:2002su} is used to relate the triple-real and the fully resolved phase space, and we apply phase space sectors~\cite{Chen:2025ojp} to avoid partial fractioning the antenna functions.
The remaining terms are needed to remove the  single- and double-unresolved limits of the $X_5^0$, which are already subtracted by $\dd\sigma^{(3)}_{5,S_{1}}+\dd\sigma^{(3)}_{5,S_{2}}$.
The five-particle antenna functions mimic the full matrix element in unresolved limits and factorize onto lower-multiplicity antenna functions as follows:
\setcounter{equation}{14}
\begin{eqnarray}
    \label{suX50} X_5^0&\xrightarrow[]{\text{single unres.}}&X_3^0\otimes X_4^0\,,\\
    \label{duX50cc} X_5^0&\xrightarrow[cc]{\text{double unres.}}&X_4^0\otimes X_3^0\,,\\
    \label{duX50acc} X_5^0&\xrightarrow[acc]{\text{double unres.}}&(X_3^0\otimes X_3^0+\mathcal{S}\otimes X_3^0)\otimes X_3^0\,.
\end{eqnarray}
Once again, the over-subtraction of single-unresolved singularities is prevented by removing the iterated $X_3^0\otimes X_3^0$ contribution from the $X_4^0$ using the second term in~\eqref{RRR:term3}.

The iterated structures in $\dd\sigma_{5,S_{3}}^{(3)}$ also remove the singular behavior of $\dd\sigma_{5,S_{1}}^{(3)}$ and $\dd\sigma_{5,S_{2}}^{(3)}$ in triple-unresolved limits. In particular,~\eqref{RRR:term2} cancels the additional double-unresolved singularities of $M_{4}^0$ in~\eqref{dsigmaS1}, while~\eqref{RRR:term2} and~\eqref{RRR:term3} subtract the single-unresolved behavior of $M_{3}^0$ originating from the terms~\eqref{RR:term1} and~\eqref{RR:term2}, respectively.
Finally, note that in $\dd\sigma_{5,S_{3}}^{(3)}$, there is no analogue of the term~\eqref{RR:term2} in $\dd\sigma_{5,S_{2}}^{(3)}$.
This is due to the fact that with up to five partons in the final state, the three unresolved emissions must occur between the same pair of hard radiators, and they are therefore always color-connected.
For higher-multiplicity processes, such as three-jet production at N$^3$LO, additional terms will be required to subtract singularities due to three unresolved partons between three or more hard radiators; for a discussion at NNLO see~\cite{Fox:2024bfp}.

\section{Calculation}
Our calculation was performed with a version of the \code{NNLOJET} Monte Carlo framework~\cite{NNLOJET:2025rno} extended for an N$^3$LO calculation.
For the N$^3$LO antenna functions $X_n^\ell$, we adapted the implementation of the relevant matrix elements in \code{EERAD3}~\cite{Gehrmann-DeRidder:2014hxk}.
The necessary helicity amplitudes for $e^+e^-\to \text{partons}$ are available in \code{NNLOJET} as part of the calculation of three-jet production at NNLO~\cite{Gehrmann:2017xfb}.
The five-parton tree-level amplitudes used in our calculation were computed in~\cite{Hagiwara:1988pp, Berends:1988yn, Falck:1988gc}, the four-parton one-loop amplitudes in~\cite{Glover:1996eh,Campbell:1997tv,Bern:1997sc} and the three-parton two-loop amplitudes in~\cite{Garland:2001tf,Garland:2002ak}.
Finally, the three-loop quark form factor was computed in~\cite{Baikov:2009bg,Gehrmann:2010ue,Lee:2010cga}.

The numerical evaluation of these matrix elements in double-unresolved configurations at one loop and single-unresolved limits at two loops required careful treatment to achieve the necessary accuracy.
In double-unresolved limits of one-loop matrix elements, we implemented a dynamical switch from double to quadruple precision for floating point operations during the evaluation of the $X_4^1$ antenna functions at phase-space points in the unresolved regions.
The switch is triggered when a two-particle invariant is lower than a given threshold, $s_{ij}/s < 10^{-5}$, where $s$ is the center-of-mass energy squared.
The two-loop quantities are expressed in terms of two-dimensional harmonic polylogarithms~\cite{Gehrmann:2000zt}.
These functions can be numerically evaluated using the library \code{tdhpl}~\cite{Gehrmann:2001jv}, which uses a representation optimized to reduce the number of arithmetic operations for points in the bulk.
For points with $s_{qg}/s < 10^{-6}$ or $s_{\bar{q}g}/s< 10^{-6}$, we substitute the amplitudes with their series expansions retaining divergent, constant and linear terms in the small invariant, obtained analytically using an algorithm~\cite{Moch:2001zr} implemented in \code{PolyLogTools}~\cite{Duhr:2019tlz}.

To validate our construction of the antenna subtraction terms at N$^3$LO, we verified that they cancel against the relevant matrix elements in all single-, double- and triple-unresolved limits at a satisfactory level for numerical integration with Monte Carlo techniques.
The complete analytic cancellation of the $\epsilon$ poles of the virtual contributions is another check on the subtraction terms.

\input{Figures/figure2.tex}

%% file: Figures/figure2.tex
\begin{figure}
    \centering
    \includegraphics[width=\linewidth]{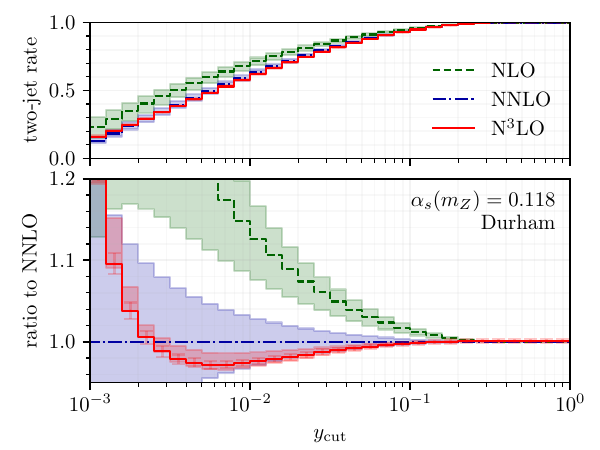}
    \caption{The two-jet production rate up to N$^3$LO, differential in the Durham jet resolution parameter $\ycut$. Error bars indicate the Monte Carlo error.
    }
    \label{fig2: jet rates}
\end{figure}

%% file: Sections/results.tex
\section{Results}
All our results are normalized by the respective Born-level cross section $\sigma_0$ for $e^+e^-\to q\bar{q}$.
As a consequence, the electroweak coupling drops out and our results are valid for both $\gamma^*$ and $Z$ as the intermediate boson.
We neglect the singlet contribution which is numerically small, as observed for example in~\cite{Dixon:1997th,Frederix:2010ne}.
We set $\sqrt{s}=m_Z=91.2$ GeV and \mbox{$\alpha_s(m_Z)=0.118$}.
However, for the observables we consider, the dependence on the value of the energy scale only enters through the running of the strong coupling.
The renormalization scale is chosen to be equal to the center-of-mass energy, $\mu_r=\sqrt{s}$, and theory uncertainties are estimated by varying it within $\mu_r\in[\sqrt{s}/2,2\sqrt{s}]$.

The first result of this Letter is the $\mathcal{O}(\alpha_s^3)$ correction to the inclusive cross section for hadron production. For five light quark flavors, we find
\begin{align}
   \sigma^{(3)} &= \sigma_0 \left(\frac{\alpha_s}{2\pi}\right)^3  (-105\pm 11)\,,
\end{align}
where the error is due to Monte Carlo integration.
This value is in agreement with the analytic result~\cite{Chetyrkin:1996ela} given by $\approx -102.14$.
We further compute the $\mathcal{O}(\alpha_s^3)$ correction to the two-jet exclusive cross section at two different values of the Durham~\cite{Catani:1991hj,Brown:1990nm,Brown:1991hx,Stirling:1991ds,Bethke:1991wk} jet reconstruction parameter:
\begin{align}
   \sigma^{(3)}_{2j}(\ycut = 0.1) &= \sigma_0 \left(\frac{\alpha_s}{2\pi}\right)^3  (-453\pm 31)\,, \\
   \sigma^{(3)}_{2j}(\ycut = 0.01) &= \sigma_0 \left(\frac{\alpha_s}{2\pi}\right)^3  (-2584\pm 40)\,. \label{eq:sig32j1dm2}
\end{align}
We verified that both values are compatible with those obtained as the difference between the inclusive cross section and the three-jet production cross section at NNLO.
The relative error is smaller for~\eqref{eq:sig32j1dm2} because it is less affected by large cancellations occurring in more inclusive setups.

\input{Figures/figure3}

As the first example of a differential distribution, we compute the two-jet production rate up to $\mathcal{O}(\alpha_s^3)$ differentially in the resolution parameter $\ycut$ of the Durham algorithm, shown in Figure~\ref{fig2: jet rates}.
We also verified that we can reproduce the result of the indirect calculation presented in Figure 2 of~\cite{Gehrmann-DeRidder:2008qsl} after accounting for a different value of $\alpha_s$.

As a second example, we consider the bin-integrated distribution
of the energy of the more energetic jet in exclusive two-jet events (in the four-momentum recombination scheme), which was calculated at NNLO in~\cite{Anastasiou:2004qd} using the JADE~\cite{JADE:1986kta}
algorithm.
We recovered the NNLO results and we computed for the first time the
$\mathcal{O}(\alpha_s^3)$ correction shown in Figure~\ref{fig3: ej1} for the Durham algorithm with $\ycut=0.1$ and $\ycut=0.01$.
For $\ycut=0.1$, we observe very good perturbative convergence with percent-level N$^3$LO corrections lying within the theoretical uncertainty of the previous order across the entire considered range.
On the other hand, for $\ycut=0.01$, the NLO and NNLO corrections quickly vanish as the energy increases and, accordingly, the N$^3$LO correction becomes comparatively large because the nominal perturbative accuracy in the tail of the distribution is lower.
Indeed, a highly energetic jet must recoil against multiple partonic emissions, which are more likely to be clustered as three or more jets for a smaller value of $\ycut$.
For both choices of $\ycut$, real radiation is suppressed close to $E_{j,1}=\sqrt{s}/2$, affecting the cancellation of IR singularities and calling for all-order resummation in order to obtain physical results, as discussed e.g. in~\cite{Mondini:2019gid}.

%% file: Figures/figure3.tex
\begin{figure*}
\centering
\begin{subfigure}{0.49\textwidth}
    \includegraphics[width=\linewidth]{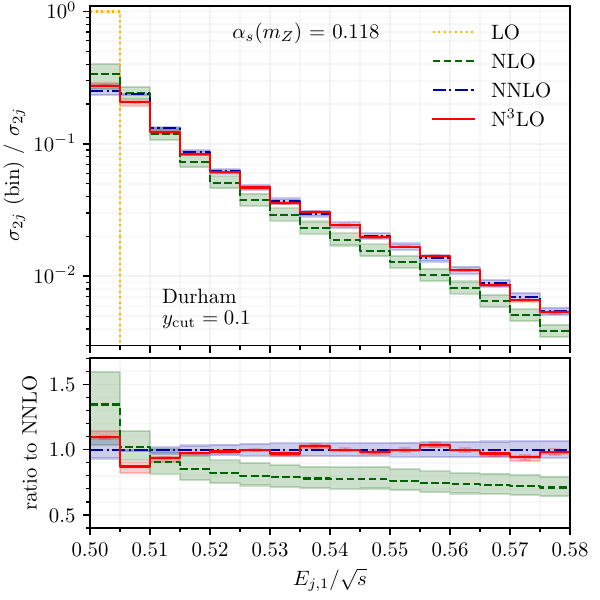}
        %\caption{
        %Durham reconstruction algorithm with $\ycut=0.1$.
        %}
    \label{fig3a: ej1 0.1}
\end{subfigure}
\hfill
\begin{subfigure}{0.49\textwidth}
    \centering
    \includegraphics[width=\linewidth]{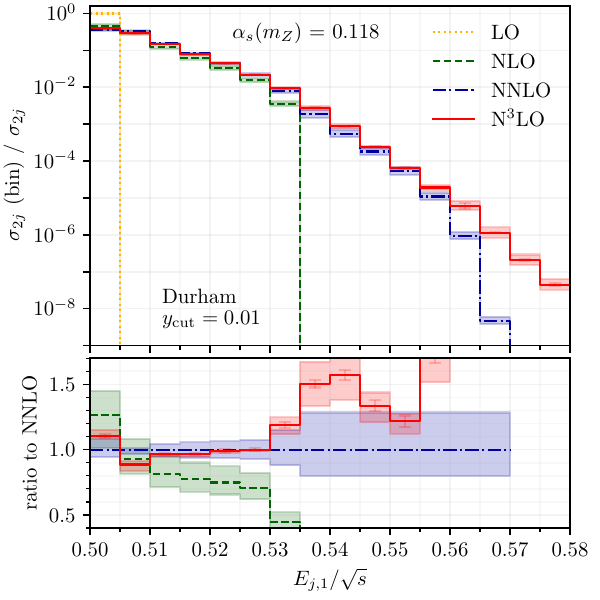}
       %\caption{Durham reconstruction algorithm with $\ycut=0.01$.}
    \label{fig3b: ej1 0.01}
\end{subfigure}
\vspace{-12pt}
\caption{
Distribution of the energy of the more energetic jet $E_{j,1}$ as a fraction of the center-of-mass energy $\sqrt{s}$ for events with two jets reconstructed with the Durham algorithm and $\ycut=0.1$ (left) or $\ycut=0.01$ (right). Error bars indicate the Monte Carlo error.
}
\label{fig3: ej1}
\end{figure*}

%% file: Sections/conclusion.tex
\section{Conclusions}
We presented the first direct fully differential calculation of jet production in electron-positron annihilation at third order in QCD.
We described the structure of the IR counterterms for the final-state radiation in $e^+e^-\to jj$ within antenna subtraction, including novel triple-unresolved limits.
As an application, we computed the production rate of two jets and the distribution of the energy of the leading jet at N$^3$LO at different values of the jet reconstruction parameter.

The future generalization of the presented N$^3$LO antenna subtraction terms to higher-multiplicity final states and hadronic initial states will likely build on the definition of optimal antenna functions~\cite{Braun-White:2023sgd,Braun-White:2023zwd,Fox:2023bma,Fox:2024bfp} and recent steps towards the integration of N$^3$LO antenna functions for initial-state radiation~\cite{Fontana:2024olm}.
Our work is an important milestone in establishing a general method for computing N$^3$LO predictions to processes involving final-state jets.

%% file: main.bbl
\begin{thebibliography}{108}
\expandafter\ifx\csname natexlab\endcsname\relax\def\natexlab#1{#1}\fi
\providecommand{\url}[1]{\texttt{#1}}
\providecommand{\href}[2]{#2}
\providecommand{\path}[1]{#1}
\providecommand{\DOIprefix}{doi:}
\providecommand{\ArXivprefix}{arXiv:}
\providecommand{\URLprefix}{URL: }
\providecommand{\Pubmedprefix}{pmid:}
\providecommand{\doi}[1]{\href{http://dx.doi.org/#1}{\path{#1}}}
\providecommand{\Pubmed}[1]{\href{pmid:#1}{\path{#1}}}
\providecommand{\bibinfo}[2]{#2}
\ifx\xfnm\relax \def\xfnm[#1]{\unskip,\space#1}\fi
%Type = Article
\bibitem[{Abada et~al.(2019a)}]{FCC:2018evy}
\bibinfo{author}{Abada, A.}, et~al. (\bibinfo{collaboration}{FCC}),
  \bibinfo{year}{2019}a.
\newblock \bibinfo{title}{{FCC-ee: The Lepton Collider}: {Future Circular
  Collider Conceptual Design Report Volume 2}}.
\newblock \bibinfo{journal}{Eur. Phys. J. ST} \bibinfo{volume}{228},
  \bibinfo{pages}{261--623}.
\newblock \DOIprefix\doi{10.1140/epjst/e2019-900045-4}.
%Type = Article
\bibitem[{Abada et~al.(2019b)}]{FCC:2018byv}
\bibinfo{author}{Abada, A.}, et~al. (\bibinfo{collaboration}{FCC}),
  \bibinfo{year}{2019}b.
\newblock \bibinfo{title}{{FCC Physics Opportunities}: {Future Circular
  Collider Conceptual Design Report Volume 1}}.
\newblock \bibinfo{journal}{Eur. Phys. J. C} \bibinfo{volume}{79},
  \bibinfo{pages}{474}.
\newblock \DOIprefix\doi{10.1140/epjc/s10052-019-6904-3}.
%Type = Article
\bibitem[{Ai et~al.(2024)}]{Ai:2024nmn}
\bibinfo{author}{Ai, X.}, et~al., \bibinfo{year}{2024}.
\newblock \bibinfo{title}{{Flavor Physics at CEPC: a General Perspective}}
  \href{http://arxiv.org/abs/2412.19743}{{\tt arXiv:2412.19743}}.
%Type = Article
\bibitem[{An et~al.(2019)}]{An:2018dwb}
\bibinfo{author}{An, F.}, et~al., \bibinfo{year}{2019}.
\newblock \bibinfo{title}{{Precision Higgs physics at the CEPC}}.
\newblock \bibinfo{journal}{Chin. Phys. C} \bibinfo{volume}{43},
  \bibinfo{pages}{043002}.
\newblock \DOIprefix\doi{10.1088/1674-1137/43/4/043002},
  \href{http://arxiv.org/abs/1810.09037}{{\tt arXiv:1810.09037}}.
%Type = Article
\bibitem[{Anastasiou et~al.(2015)Anastasiou, Duhr, Dulat, Herzog and
  Mistlberger}]{Anastasiou:2015vya}
\bibinfo{author}{Anastasiou, C.}, \bibinfo{author}{Duhr, C.},
  \bibinfo{author}{Dulat, F.}, \bibinfo{author}{Herzog, F.},
  \bibinfo{author}{Mistlberger, B.}, \bibinfo{year}{2015}.
\newblock \bibinfo{title}{{Higgs Boson Gluon-Fusion Production in QCD at Three
  Loops}}.
\newblock \bibinfo{journal}{Phys. Rev. Lett.} \bibinfo{volume}{114},
  \bibinfo{pages}{212001}.
\newblock \DOIprefix\doi{10.1103/PhysRevLett.114.212001},
  \href{http://arxiv.org/abs/1503.06056}{{\tt arXiv:1503.06056}}.
%Type = Article
\bibitem[{Anastasiou et~al.(2004)Anastasiou, Melnikov and
  Petriello}]{Anastasiou:2004qd}
\bibinfo{author}{Anastasiou, C.}, \bibinfo{author}{Melnikov, K.},
  \bibinfo{author}{Petriello, F.}, \bibinfo{year}{2004}.
\newblock \bibinfo{title}{{Real radiation at NNLO: $e^+ e^- \to$ 2 jets through
  $\mathcal{O}(\alpha_s^2)$}}.
\newblock \bibinfo{journal}{Phys. Rev. Lett.} \bibinfo{volume}{93},
  \bibinfo{pages}{032002}.
\newblock \DOIprefix\doi{10.1103/PhysRevLett.93.032002},
  \href{http://arxiv.org/abs/hep-ph/0402280}{{\tt arXiv:hep-ph/0402280}}.
%Type = Article
\bibitem[{Baglio et~al.(2022)Baglio, Duhr, Mistlberger and
  Szafron}]{Baglio:2022wzu}
\bibinfo{author}{Baglio, J.}, \bibinfo{author}{Duhr, C.},
  \bibinfo{author}{Mistlberger, B.}, \bibinfo{author}{Szafron, R.},
  \bibinfo{year}{2022}.
\newblock \bibinfo{title}{{Inclusive production cross sections at N$^{3}$LO}}.
\newblock \bibinfo{journal}{JHEP} \bibinfo{volume}{12}, \bibinfo{pages}{066}.
\newblock \DOIprefix\doi{10.1007/JHEP12(2022)066},
  \href{http://arxiv.org/abs/2209.06138}{{\tt arXiv:2209.06138}}.
%Type = Article
\bibitem[{Baikov et~al.(2009)Baikov, Chetyrkin, Smirnov, Smirnov and
  Steinhauser}]{Baikov:2009bg}
\bibinfo{author}{Baikov, P.A.}, \bibinfo{author}{Chetyrkin, K.G.},
  \bibinfo{author}{Smirnov, A.V.}, \bibinfo{author}{Smirnov, V.A.},
  \bibinfo{author}{Steinhauser, M.}, \bibinfo{year}{2009}.
\newblock \bibinfo{title}{{Quark and gluon form factors to three loops}}.
\newblock \bibinfo{journal}{Phys. Rev. Lett.} \bibinfo{volume}{102},
  \bibinfo{pages}{212002}.
\newblock \DOIprefix\doi{10.1103/PhysRevLett.102.212002},
  \href{http://arxiv.org/abs/0902.3519}{{\tt arXiv:0902.3519}}.
%Type = Article
\bibitem[{Bartel et~al.(1986)}]{JADE:1986kta}
\bibinfo{author}{Bartel, W.}, et~al. (\bibinfo{collaboration}{JADE}),
  \bibinfo{year}{1986}.
\newblock \bibinfo{title}{{Experimental Studies on Multi-Jet Production in $e^+
  e^-$ Annihilation at PETRA Energies}}.
\newblock \bibinfo{journal}{Z. Phys. C} \bibinfo{volume}{33},
  \bibinfo{pages}{23}.
\newblock \DOIprefix\doi{10.1007/BF01410449}.
%Type = Article
\bibitem[{Benedikt et~al.(2025)}]{FCC:2025lpp}
\bibinfo{author}{Benedikt, M.}, et~al. (\bibinfo{collaboration}{FCC}),
  \bibinfo{year}{2025}.
\newblock \bibinfo{title}{{Future Circular Collider Feasibility Study Report:
  Volume 1, Physics, Experiments, Detectors}}
  \DOIprefix\doi{10.17181/CERN.9DKX.TDH9},
  \href{http://arxiv.org/abs/2505.00272}{{\tt arXiv:2505.00272}}.
%Type = Article
\bibitem[{Berends et~al.(1989)Berends, Giele and Kuijf}]{Berends:1988yn}
\bibinfo{author}{Berends, F.A.}, \bibinfo{author}{Giele, W.T.},
  \bibinfo{author}{Kuijf, H.}, \bibinfo{year}{1989}.
\newblock \bibinfo{title}{{Exact Expressions for Processes Involving a Vector
  Boson and Up to Five Partons}}.
\newblock \bibinfo{journal}{Nucl. Phys. B} \bibinfo{volume}{321},
  \bibinfo{pages}{39--82}.
\newblock \DOIprefix\doi{10.1016/0550-3213(89)90242-3}.
%Type = Article
\bibitem[{Bern et~al.(1998)Bern, Dixon and Kosower}]{Bern:1997sc}
\bibinfo{author}{Bern, Z.}, \bibinfo{author}{Dixon, L.J.},
  \bibinfo{author}{Kosower, D.A.}, \bibinfo{year}{1998}.
\newblock \bibinfo{title}{{One loop amplitudes for e$^{+}$e$^{-}$ to four
  partons}}.
\newblock \bibinfo{journal}{Nucl. Phys. B} \bibinfo{volume}{513},
  \bibinfo{pages}{3--86}.
\newblock \DOIprefix\doi{10.1016/S0550-3213(97)00703-7},
  \href{http://arxiv.org/abs/hep-ph/9708239}{{\tt arXiv:hep-ph/9708239}}.
%Type = Article
\bibitem[{Bertolotti et~al.(2023)Bertolotti, Magnea, Pelliccioli, Ratti,
  Signorile-Signorile, Torrielli and Uccirati}]{Bertolotti:2022aih}
\bibinfo{author}{Bertolotti, G.}, \bibinfo{author}{Magnea, L.},
  \bibinfo{author}{Pelliccioli, G.}, \bibinfo{author}{Ratti, A.},
  \bibinfo{author}{Signorile-Signorile, C.}, \bibinfo{author}{Torrielli, P.},
  \bibinfo{author}{Uccirati, S.}, \bibinfo{year}{2023}.
\newblock \bibinfo{title}{{NNLO subtraction for any massless final state: a
  complete analytic expression}}.
\newblock \bibinfo{journal}{JHEP} \bibinfo{volume}{07}, \bibinfo{pages}{140}.
\newblock \DOIprefix\doi{10.1007/JHEP07(2023)140},
  \href{http://arxiv.org/abs/2212.11190}{{\tt arXiv:2212.11190}}.
  \bibinfo{note}{[Erratum: JHEP 05, 019 (2024)]}.
%Type = Article
\bibitem[{Bethke et~al.(1992)Bethke, Kunszt, Soper and
  Stirling}]{Bethke:1991wk}
\bibinfo{author}{Bethke, S.}, \bibinfo{author}{Kunszt, Z.},
  \bibinfo{author}{Soper, D.E.}, \bibinfo{author}{Stirling, W.J.},
  \bibinfo{year}{1992}.
\newblock \bibinfo{title}{{New jet cluster algorithms: Next-to-leading order
  QCD and hadronization corrections}}.
\newblock \bibinfo{journal}{Nucl. Phys. B} \bibinfo{volume}{370},
  \bibinfo{pages}{310--334}.
\newblock \DOIprefix\doi{10.1016/0550-3213(92)90289-N}.
  \bibinfo{note}{[Erratum: Nucl.Phys.B 523, 681--681 (1998)]}.
%Type = Article
\bibitem[{Billis et~al.(2021)Billis, Dehnadi, Ebert, Michel and
  Tackmann}]{Billis:2021ecs}
\bibinfo{author}{Billis, G.}, \bibinfo{author}{Dehnadi, B.},
  \bibinfo{author}{Ebert, M.A.}, \bibinfo{author}{Michel, J.K.L.},
  \bibinfo{author}{Tackmann, F.J.}, \bibinfo{year}{2021}.
\newblock \bibinfo{title}{{Higgs pT Spectrum and Total Cross Section with
  Fiducial Cuts at Third Resummed and Fixed Order in QCD}}.
\newblock \bibinfo{journal}{Phys. Rev. Lett.} \bibinfo{volume}{127},
  \bibinfo{pages}{072001}.
\newblock \DOIprefix\doi{10.1103/PhysRevLett.127.072001},
  \href{http://arxiv.org/abs/2102.08039}{{\tt arXiv:2102.08039}}.
%Type = Article
\bibitem[{Bonino et~al.(2024)Bonino, Gehrmann, Marcoli, Sch\"urmann and
  Stagnitto}]{Bonino:2024adk}
\bibinfo{author}{Bonino, L.}, \bibinfo{author}{Gehrmann, T.},
  \bibinfo{author}{Marcoli, M.}, \bibinfo{author}{Sch\"urmann, R.},
  \bibinfo{author}{Stagnitto, G.}, \bibinfo{year}{2024}.
\newblock \bibinfo{title}{{Antenna subtraction for processes with identified
  particles at hadron colliders}}.
\newblock \bibinfo{journal}{JHEP} \bibinfo{volume}{08}, \bibinfo{pages}{073}.
\newblock \DOIprefix\doi{10.1007/JHEP08(2024)073},
  \href{http://arxiv.org/abs/2406.09925}{{\tt arXiv:2406.09925}}.
%Type = Article
\bibitem[{Boughezal et~al.(2012)Boughezal, Melnikov and
  Petriello}]{Boughezal:2011jf}
\bibinfo{author}{Boughezal, R.}, \bibinfo{author}{Melnikov, K.},
  \bibinfo{author}{Petriello, F.}, \bibinfo{year}{2012}.
\newblock \bibinfo{title}{{A subtraction scheme for NNLO computations}}.
\newblock \bibinfo{journal}{Phys. Rev. D} \bibinfo{volume}{85},
  \bibinfo{pages}{034025}.
\newblock \DOIprefix\doi{10.1103/PhysRevD.85.034025},
  \href{http://arxiv.org/abs/1111.7041}{{\tt arXiv:1111.7041}}.
%Type = Article
\bibitem[{Braun-White et~al.(2023a)Braun-White, Glover and
  Preuss}]{Braun-White:2023zwd}
\bibinfo{author}{Braun-White, O.}, \bibinfo{author}{Glover, N.},
  \bibinfo{author}{Preuss, C.T.}, \bibinfo{year}{2023}a.
\newblock \bibinfo{title}{{A general algorithm to build mixed real and virtual
  antenna functions for higher-order calculations}}.
\newblock \bibinfo{journal}{JHEP} \bibinfo{volume}{11}, \bibinfo{pages}{179}.
\newblock \DOIprefix\doi{10.1007/JHEP11(2023)179},
  \href{http://arxiv.org/abs/2307.14999}{{\tt arXiv:2307.14999}}.
%Type = Article
\bibitem[{Braun-White et~al.(2023b)Braun-White, Glover and
  Preuss}]{Braun-White:2023sgd}
\bibinfo{author}{Braun-White, O.}, \bibinfo{author}{Glover, N.},
  \bibinfo{author}{Preuss, C.T.}, \bibinfo{year}{2023}b.
\newblock \bibinfo{title}{{A general algorithm to build real-radiation antenna
  functions for higher-order calculations}}.
\newblock \bibinfo{journal}{JHEP} \bibinfo{volume}{06}, \bibinfo{pages}{065}.
\newblock \DOIprefix\doi{10.1007/JHEP06(2023)065},
  \href{http://arxiv.org/abs/2302.12787}{{\tt arXiv:2302.12787}}.
%Type = Article
\bibitem[{Brown and Stirling(1990)}]{Brown:1990nm}
\bibinfo{author}{Brown, N.}, \bibinfo{author}{Stirling, W.J.},
  \bibinfo{year}{1990}.
\newblock \bibinfo{title}{{Jet cross-sections at leading double logarithm in
  $e^+ e^-$ annihilation}}.
\newblock \bibinfo{journal}{Phys. Lett. B} \bibinfo{volume}{252},
  \bibinfo{pages}{657--662}.
\newblock \DOIprefix\doi{10.1016/0370-2693(90)90502-W}.
%Type = Article
\bibitem[{Brown and Stirling(1992)}]{Brown:1991hx}
\bibinfo{author}{Brown, N.}, \bibinfo{author}{Stirling, W.J.},
  \bibinfo{year}{1992}.
\newblock \bibinfo{title}{{Finding jets and summing soft gluons: A New
  algorithm}}.
\newblock \bibinfo{journal}{Z. Phys. C} \bibinfo{volume}{53},
  \bibinfo{pages}{629--636}.
\newblock \DOIprefix\doi{10.1007/BF01559740}.
%Type = Article
\bibitem[{Cacciari et~al.(2015)Cacciari, Dreyer, Karlberg, Salam and
  Zanderighi}]{Cacciari:2015jma}
\bibinfo{author}{Cacciari, M.}, \bibinfo{author}{Dreyer, F.A.},
  \bibinfo{author}{Karlberg, A.}, \bibinfo{author}{Salam, G.P.},
  \bibinfo{author}{Zanderighi, G.}, \bibinfo{year}{2015}.
\newblock \bibinfo{title}{{Fully Differential Vector-Boson-Fusion Higgs
  Production at Next-to-Next-to-Leading Order}}.
\newblock \bibinfo{journal}{Phys. Rev. Lett.} \bibinfo{volume}{115},
  \bibinfo{pages}{082002}.
\newblock \DOIprefix\doi{10.1103/PhysRevLett.115.082002},
  \href{http://arxiv.org/abs/1506.02660}{{\tt arXiv:1506.02660}}.
  \bibinfo{note}{[Erratum: Phys.Rev.Lett. 120, 139901 (2018)]}.
%Type = Article
\bibitem[{Campbell and Neumann(2023)}]{Campbell:2023lcy}
\bibinfo{author}{Campbell, J.}, \bibinfo{author}{Neumann, T.},
  \bibinfo{year}{2023}.
\newblock \bibinfo{title}{{Third order QCD predictions for fiducial W-boson
  production}}.
\newblock \bibinfo{journal}{JHEP} \bibinfo{volume}{11}, \bibinfo{pages}{127}.
\newblock \DOIprefix\doi{10.1007/JHEP11(2023)127},
  \href{http://arxiv.org/abs/2308.15382}{{\tt arXiv:2308.15382}}.
%Type = Article
\bibitem[{Campbell et~al.(1999)Campbell, Cullen and Glover}]{Campbell:1998nn}
\bibinfo{author}{Campbell, J.M.}, \bibinfo{author}{Cullen, M.A.},
  \bibinfo{author}{Glover, E.W.N.}, \bibinfo{year}{1999}.
\newblock \bibinfo{title}{{Four jet event shapes in electron - positron
  annihilation}}.
\newblock \bibinfo{journal}{Eur. Phys. J. C} \bibinfo{volume}{9},
  \bibinfo{pages}{245--265}.
\newblock \DOIprefix\doi{10.1007/s100529900034},
  \href{http://arxiv.org/abs/hep-ph/9809429}{{\tt arXiv:hep-ph/9809429}}.
%Type = Article
\bibitem[{Campbell et~al.(1997)Campbell, Glover and Miller}]{Campbell:1997tv}
\bibinfo{author}{Campbell, J.M.}, \bibinfo{author}{Glover, E.W.N.},
  \bibinfo{author}{Miller, D.J.}, \bibinfo{year}{1997}.
\newblock \bibinfo{title}{{The One loop QCD corrections for $\gamma^* \to q
  \bar{q} gg$}}.
\newblock \bibinfo{journal}{Phys. Lett. B} \bibinfo{volume}{409},
  \bibinfo{pages}{503--508}.
\newblock \DOIprefix\doi{10.1016/S0370-2693(97)00909-X},
  \href{http://arxiv.org/abs/hep-ph/9706297}{{\tt arXiv:hep-ph/9706297}}.
%Type = Inproceedings
\bibitem[{Caola et~al.(2022)Caola, Chen, Duhr, Liu, Mistlberger, Petriello,
  Vita and Weinzierl}]{Caola:2022ayt}
\bibinfo{author}{Caola, F.}, \bibinfo{author}{Chen, W.}, \bibinfo{author}{Duhr,
  C.}, \bibinfo{author}{Liu, X.}, \bibinfo{author}{Mistlberger, B.},
  \bibinfo{author}{Petriello, F.}, \bibinfo{author}{Vita, G.},
  \bibinfo{author}{Weinzierl, S.}, \bibinfo{year}{2022}.
\newblock \bibinfo{title}{{The Path forward to N$^3$LO}}, in:
  \bibinfo{booktitle}{{Snowmass 2021}}.
\newblock \href{http://arxiv.org/abs/2203.06730}{{\tt arXiv:2203.06730}}.
%Type = Article
\bibitem[{Caola et~al.(2017)Caola, Melnikov and R\"ontsch}]{Caola:2017dug}
\bibinfo{author}{Caola, F.}, \bibinfo{author}{Melnikov, K.},
  \bibinfo{author}{R\"ontsch, R.}, \bibinfo{year}{2017}.
\newblock \bibinfo{title}{{Nested soft-collinear subtractions in NNLO QCD
  computations}}.
\newblock \bibinfo{journal}{Eur. Phys. J. C} \bibinfo{volume}{77},
  \bibinfo{pages}{248}.
\newblock \DOIprefix\doi{10.1140/epjc/s10052-017-4774-0},
  \href{http://arxiv.org/abs/1702.01352}{{\tt arXiv:1702.01352}}.
%Type = Article
\bibitem[{Catani et~al.(2023)Catani, Cieri, Colferai and
  Coradeschi}]{Catani:2022hkb}
\bibinfo{author}{Catani, S.}, \bibinfo{author}{Cieri, L.},
  \bibinfo{author}{Colferai, D.}, \bibinfo{author}{Coradeschi, F.},
  \bibinfo{year}{2023}.
\newblock \bibinfo{title}{{Soft gluon\textendash{}quark\textendash{}antiquark
  emission in QCD hard scattering}}.
\newblock \bibinfo{journal}{Eur. Phys. J. C} \bibinfo{volume}{83},
  \bibinfo{pages}{38}.
\newblock \DOIprefix\doi{10.1140/epjc/s10052-022-11141-y},
  \href{http://arxiv.org/abs/2210.09397}{{\tt arXiv:2210.09397}}.
%Type = Article
\bibitem[{Catani et~al.(2020)Catani, Colferai and Torrini}]{Catani:2019nqv}
\bibinfo{author}{Catani, S.}, \bibinfo{author}{Colferai, D.},
  \bibinfo{author}{Torrini, A.}, \bibinfo{year}{2020}.
\newblock \bibinfo{title}{{Triple (and quadruple) soft-gluon radiation in QCD
  hard scattering}}.
\newblock \bibinfo{journal}{JHEP} \bibinfo{volume}{01}, \bibinfo{pages}{118}.
\newblock \DOIprefix\doi{10.1007/JHEP01(2020)118},
  \href{http://arxiv.org/abs/1908.01616}{{\tt arXiv:1908.01616}}.
%Type = Article
\bibitem[{Catani et~al.(1991)Catani, Dokshitzer, Olsson, Turnock and
  Webber}]{Catani:1991hj}
\bibinfo{author}{Catani, S.}, \bibinfo{author}{Dokshitzer, Y.L.},
  \bibinfo{author}{Olsson, M.}, \bibinfo{author}{Turnock, G.},
  \bibinfo{author}{Webber, B.R.}, \bibinfo{year}{1991}.
\newblock \bibinfo{title}{{New clustering algorithm for multi - jet
  cross-sections in $e^+ e^-$ annihilation}}.
\newblock \bibinfo{journal}{Phys. Lett. B} \bibinfo{volume}{269},
  \bibinfo{pages}{432--438}.
\newblock \DOIprefix\doi{10.1016/0370-2693(91)90196-W}.
%Type = Article
\bibitem[{Catani and Grazzini(2007)}]{Catani:2007vq}
\bibinfo{author}{Catani, S.}, \bibinfo{author}{Grazzini, M.},
  \bibinfo{year}{2007}.
\newblock \bibinfo{title}{{An NNLO subtraction formalism in hadron collisions
  and its application to Higgs boson production at the LHC}}.
\newblock \bibinfo{journal}{Phys. Rev. Lett.} \bibinfo{volume}{98},
  \bibinfo{pages}{222002}.
\newblock \DOIprefix\doi{10.1103/PhysRevLett.98.222002},
  \href{http://arxiv.org/abs/hep-ph/0703012}{{\tt arXiv:hep-ph/0703012}}.
%Type = Article
\bibitem[{Catani and Seymour(1997)}]{Catani:1996vz}
\bibinfo{author}{Catani, S.}, \bibinfo{author}{Seymour, M.H.},
  \bibinfo{year}{1997}.
\newblock \bibinfo{title}{{A General algorithm for calculating jet
  cross-sections in NLO QCD}}.
\newblock \bibinfo{journal}{Nucl. Phys. B} \bibinfo{volume}{485},
  \bibinfo{pages}{291--419}.
\newblock \DOIprefix\doi{10.1016/S0550-3213(96)00589-5},
  \href{http://arxiv.org/abs/hep-ph/9605323}{{\tt arXiv:hep-ph/9605323}}.
  \bibinfo{note}{[Erratum: Nucl.Phys.B 510, 503--504 (1998)]}.
%Type = Article
\bibitem[{Chen et~al.(2020a)Chen, Li, Shao and Wang}]{Chen:2019lzz}
\bibinfo{author}{Chen, L.B.}, \bibinfo{author}{Li, H.T.},
  \bibinfo{author}{Shao, H.S.}, \bibinfo{author}{Wang, J.},
  \bibinfo{year}{2020}a.
\newblock \bibinfo{title}{{Higgs boson pair production via gluon fusion at
  N$^3$LO in QCD}}.
\newblock \bibinfo{journal}{Phys. Lett. B} \bibinfo{volume}{803},
  \bibinfo{pages}{135292}.
\newblock \DOIprefix\doi{10.1016/j.physletb.2020.135292},
  \href{http://arxiv.org/abs/1909.06808}{{\tt arXiv:1909.06808}}.
%Type = Article
\bibitem[{Chen et~al.(2020b)Chen, Li, Shao and Wang}]{Chen:2019fhs}
\bibinfo{author}{Chen, L.B.}, \bibinfo{author}{Li, H.T.},
  \bibinfo{author}{Shao, H.S.}, \bibinfo{author}{Wang, J.},
  \bibinfo{year}{2020}b.
\newblock \bibinfo{title}{{The gluon-fusion production of Higgs boson pair:
  N$^3$LO QCD corrections and top-quark mass effects}}.
\newblock \bibinfo{journal}{JHEP} \bibinfo{volume}{03}, \bibinfo{pages}{072}.
\newblock \DOIprefix\doi{10.1007/JHEP03(2020)072},
  \href{http://arxiv.org/abs/1912.13001}{{\tt arXiv:1912.13001}}.
%Type = Article
\bibitem[{Chen et~al.(2021)Chen, Gehrmann, Glover, Huss, Mistlberger and
  Pelloni}]{Chen:2021isd}
\bibinfo{author}{Chen, X.}, \bibinfo{author}{Gehrmann, T.},
  \bibinfo{author}{Glover, E.W.N.}, \bibinfo{author}{Huss, A.},
  \bibinfo{author}{Mistlberger, B.}, \bibinfo{author}{Pelloni, A.},
  \bibinfo{year}{2021}.
\newblock \bibinfo{title}{{Fully Differential Higgs Boson Production to Third
  Order in QCD}}.
\newblock \bibinfo{journal}{Phys. Rev. Lett.} \bibinfo{volume}{127},
  \bibinfo{pages}{072002}.
\newblock \DOIprefix\doi{10.1103/PhysRevLett.127.072002},
  \href{http://arxiv.org/abs/2102.07607}{{\tt arXiv:2102.07607}}.
%Type = Article
\bibitem[{Chen et~al.(2022a)Chen, Gehrmann, Glover, Huss, Monni, Re, Rottoli
  and Torrielli}]{Chen:2022cgv}
\bibinfo{author}{Chen, X.}, \bibinfo{author}{Gehrmann, T.},
  \bibinfo{author}{Glover, E.W.N.}, \bibinfo{author}{Huss, A.},
  \bibinfo{author}{Monni, P.F.}, \bibinfo{author}{Re, E.},
  \bibinfo{author}{Rottoli, L.}, \bibinfo{author}{Torrielli, P.},
  \bibinfo{year}{2022}a.
\newblock \bibinfo{title}{{Third-Order Fiducial Predictions for Drell-Yan
  Production at the LHC}}.
\newblock \bibinfo{journal}{Phys. Rev. Lett.} \bibinfo{volume}{128},
  \bibinfo{pages}{252001}.
\newblock \DOIprefix\doi{10.1103/PhysRevLett.128.252001},
  \href{http://arxiv.org/abs/2203.01565}{{\tt arXiv:2203.01565}}.
%Type = Article
\bibitem[{Chen et~al.(2022b)Chen, Gehrmann, Glover, Huss, Yang and
  Zhu}]{Chen:2021vtu}
\bibinfo{author}{Chen, X.}, \bibinfo{author}{Gehrmann, T.},
  \bibinfo{author}{Glover, N.}, \bibinfo{author}{Huss, A.},
  \bibinfo{author}{Yang, T.Z.}, \bibinfo{author}{Zhu, H.X.},
  \bibinfo{year}{2022}b.
\newblock \bibinfo{title}{{Dilepton Rapidity Distribution in Drell-Yan
  Production to Third Order in QCD}}.
\newblock \bibinfo{journal}{Phys. Rev. Lett.} \bibinfo{volume}{128},
  \bibinfo{pages}{052001}.
\newblock \DOIprefix\doi{10.1103/PhysRevLett.128.052001},
  \href{http://arxiv.org/abs/2107.09085}{{\tt arXiv:2107.09085}}.
%Type = Article
\bibitem[{Chen et~al.(2023a)Chen, Gehrmann, Glover, Huss, Yang and
  Zhu}]{Chen:2022lwc}
\bibinfo{author}{Chen, X.}, \bibinfo{author}{Gehrmann, T.},
  \bibinfo{author}{Glover, N.}, \bibinfo{author}{Huss, A.},
  \bibinfo{author}{Yang, T.Z.}, \bibinfo{author}{Zhu, H.X.},
  \bibinfo{year}{2023}a.
\newblock \bibinfo{title}{{Transverse mass distribution and charge asymmetry in
  W boson production to third order in QCD}}.
\newblock \bibinfo{journal}{Phys. Lett. B} \bibinfo{volume}{840},
  \bibinfo{pages}{137876}.
\newblock \DOIprefix\doi{10.1016/j.physletb.2023.137876},
  \href{http://arxiv.org/abs/2205.11426}{{\tt arXiv:2205.11426}}.
%Type = Article
\bibitem[{Chen et~al.(2023b)Chen, Jakub\v{c}\'\i{}k, Marcoli and
  Stagnitto}]{Chen:2023egxNeutralino}
\bibinfo{author}{Chen, X.}, \bibinfo{author}{Jakub\v{c}\'\i{}k, P.},
  \bibinfo{author}{Marcoli, M.}, \bibinfo{author}{Stagnitto, G.},
  \bibinfo{year}{2023}b.
\newblock \bibinfo{title}{{Radiation from a gluon-gluino colour-singlet dipole
  at N$^{3}$LO}}.
\newblock \bibinfo{journal}{JHEP} \bibinfo{volume}{12}, \bibinfo{pages}{198}.
\newblock \DOIprefix\doi{10.1007/JHEP12(2023)198},
  \href{http://arxiv.org/abs/2310.13062}{{\tt arXiv:2310.13062}}.
%Type = Article
\bibitem[{Chen et~al.(2023c)Chen, Jakub\v{c}\'\i{}k, Marcoli and
  Stagnitto}]{Chen:2023fbaHiggs}
\bibinfo{author}{Chen, X.}, \bibinfo{author}{Jakub\v{c}\'\i{}k, P.},
  \bibinfo{author}{Marcoli, M.}, \bibinfo{author}{Stagnitto, G.},
  \bibinfo{year}{2023}c.
\newblock \bibinfo{title}{{The parton-level structure of Higgs decays to
  hadrons at N$^{3}$LO}}.
\newblock \bibinfo{journal}{JHEP} \bibinfo{volume}{06}, \bibinfo{pages}{185}.
\newblock \DOIprefix\doi{10.1007/JHEP06(2023)185},
  \href{http://arxiv.org/abs/2304.11180}{{\tt arXiv:2304.11180}}.
%Type = Article
\bibitem[{Chen and Marcoli(2025)}]{Chen:2025ojp}
\bibinfo{author}{Chen, X.}, \bibinfo{author}{Marcoli, M.},
  \bibinfo{year}{2025}.
\newblock \bibinfo{title}{{Phase-space sectors for ordered momentum mappings in
  local subtraction up to N$^3$LO}} \href{http://arxiv.org/abs/2507.12537}{{\tt
  arXiv:2507.12537}}.
%Type = Inproceedings
\bibitem[{Cheng et~al.(2022)}]{CEPCPhysicsStudyGroup:2022uwl}
\bibinfo{author}{Cheng, H.}, et~al. (\bibinfo{collaboration}{CEPC Physics Study
  Group}), \bibinfo{year}{2022}.
\newblock \bibinfo{title}{{The Physics potential of the CEPC. Prepared for the
  US Snowmass Community Planning Exercise (Snowmass 2021)}}, in:
  \bibinfo{booktitle}{{Snowmass 2021}}.
\newblock \href{http://arxiv.org/abs/2205.08553}{{\tt arXiv:2205.08553}}.
%Type = Article
\bibitem[{Chetyrkin et~al.(1996)Chetyrkin, Kuhn and
  Kwiatkowski}]{Chetyrkin:1996ela}
\bibinfo{author}{Chetyrkin, K.G.}, \bibinfo{author}{Kuhn, J.H.},
  \bibinfo{author}{Kwiatkowski, A.}, \bibinfo{year}{1996}.
\newblock \bibinfo{title}{{QCD corrections to the $e^{+} e^{-}$ cross-section
  and the $Z$ boson decay rate}}.
\newblock \bibinfo{journal}{Phys. Rept.} \bibinfo{volume}{277},
  \bibinfo{pages}{189--281}.
\newblock \DOIprefix\doi{10.1016/S0370-1573(96)00012-9},
  \href{http://arxiv.org/abs/hep-ph/9503396}{{\tt arXiv:hep-ph/9503396}}.
%Type = Article
\bibitem[{Cieri et~al.(2019)Cieri, Chen, Gehrmann, Glover and
  Huss}]{Cieri:2018oms}
\bibinfo{author}{Cieri, L.}, \bibinfo{author}{Chen, X.},
  \bibinfo{author}{Gehrmann, T.}, \bibinfo{author}{Glover, E.W.N.},
  \bibinfo{author}{Huss, A.}, \bibinfo{year}{2019}.
\newblock \bibinfo{title}{{Higgs boson production at the LHC using the $q_T$
  subtraction formalism at N$^3$LO QCD}}.
\newblock \bibinfo{journal}{JHEP} \bibinfo{volume}{02}, \bibinfo{pages}{096}.
\newblock \DOIprefix\doi{10.1007/JHEP02(2019)096},
  \href{http://arxiv.org/abs/1807.11501}{{\tt arXiv:1807.11501}}.
%Type = Article
\bibitem[{Currie et~al.(2013)Currie, Glover and Wells}]{Currie:2013vh}
\bibinfo{author}{Currie, J.}, \bibinfo{author}{Glover, E.W.N.},
  \bibinfo{author}{Wells, S.}, \bibinfo{year}{2013}.
\newblock \bibinfo{title}{{Infrared Structure at NNLO Using Antenna
  Subtraction}}.
\newblock \bibinfo{journal}{JHEP} \bibinfo{volume}{04}, \bibinfo{pages}{066}.
\newblock \DOIprefix\doi{10.1007/JHEP04(2013)066},
  \href{http://arxiv.org/abs/1301.4693}{{\tt arXiv:1301.4693}}.
%Type = Article
\bibitem[{Czakon(2010)}]{Czakon:2010td}
\bibinfo{author}{Czakon, M.}, \bibinfo{year}{2010}.
\newblock \bibinfo{title}{{A novel subtraction scheme for double-real radiation
  at NNLO}}.
\newblock \bibinfo{journal}{Phys. Lett. B} \bibinfo{volume}{693},
  \bibinfo{pages}{259--268}.
\newblock \DOIprefix\doi{10.1016/j.physletb.2010.08.036},
  \href{http://arxiv.org/abs/1005.0274}{{\tt arXiv:1005.0274}}.
%Type = Article
\bibitem[{Czakon and Heymes(2014)}]{Czakon:2014oma}
\bibinfo{author}{Czakon, M.}, \bibinfo{author}{Heymes, D.},
  \bibinfo{year}{2014}.
\newblock \bibinfo{title}{{Four-dimensional formulation of the sector-improved
  residue subtraction scheme}}.
\newblock \bibinfo{journal}{Nucl. Phys. B} \bibinfo{volume}{890},
  \bibinfo{pages}{152--227}.
\newblock \DOIprefix\doi{10.1016/j.nuclphysb.2014.11.006},
  \href{http://arxiv.org/abs/1408.2500}{{\tt arXiv:1408.2500}}.
%Type = Article
\bibitem[{Del~Duca et~al.(2020a)Del~Duca, Duhr, Haindl, Lazopoulos and
  Michel}]{DelDuca:2020vst}
\bibinfo{author}{Del~Duca, V.}, \bibinfo{author}{Duhr, C.},
  \bibinfo{author}{Haindl, R.}, \bibinfo{author}{Lazopoulos, A.},
  \bibinfo{author}{Michel, M.}, \bibinfo{year}{2020}a.
\newblock \bibinfo{title}{{Tree-level splitting amplitudes for a gluon into
  four collinear partons}}.
\newblock \bibinfo{journal}{JHEP} \bibinfo{volume}{10}, \bibinfo{pages}{093}.
\newblock \DOIprefix\doi{10.1007/JHEP10(2020)093},
  \href{http://arxiv.org/abs/2007.05345}{{\tt arXiv:2007.05345}}.
%Type = Article
\bibitem[{Del~Duca et~al.(2020b)Del~Duca, Duhr, Haindl, Lazopoulos and
  Michel}]{DelDuca:2019ggv}
\bibinfo{author}{Del~Duca, V.}, \bibinfo{author}{Duhr, C.},
  \bibinfo{author}{Haindl, R.}, \bibinfo{author}{Lazopoulos, A.},
  \bibinfo{author}{Michel, M.}, \bibinfo{year}{2020}b.
\newblock \bibinfo{title}{{Tree-level splitting amplitudes for a quark into
  four collinear partons}}.
\newblock \bibinfo{journal}{JHEP} \bibinfo{volume}{02}, \bibinfo{pages}{189}.
\newblock \DOIprefix\doi{10.1007/JHEP02(2020)189},
  \href{http://arxiv.org/abs/1912.06425}{{\tt arXiv:1912.06425}}.
%Type = Article
\bibitem[{Del~Duca et~al.(2023)Del~Duca, Duhr, Haindl and
  Liu}]{DelDuca:2022noh}
\bibinfo{author}{Del~Duca, V.}, \bibinfo{author}{Duhr, C.},
  \bibinfo{author}{Haindl, R.}, \bibinfo{author}{Liu, Z.},
  \bibinfo{year}{2023}.
\newblock \bibinfo{title}{{Tree-level soft emission of a quark pair in
  association with a gluon}}.
\newblock \bibinfo{journal}{JHEP} \bibinfo{volume}{01}, \bibinfo{pages}{040}.
\newblock \DOIprefix\doi{10.1007/JHEP01(2023)040},
  \href{http://arxiv.org/abs/2206.01584}{{\tt arXiv:2206.01584}}.
%Type = Article
\bibitem[{Del~Duca et~al.(2016)Del~Duca, Duhr, Kardos, Somogyi, Szor,
  {Tr\'ocs\'anyi} and {Tulip\'ant}}]{DelDuca:2016ily}
\bibinfo{author}{Del~Duca, V.}, \bibinfo{author}{Duhr, C.},
  \bibinfo{author}{Kardos, A.}, \bibinfo{author}{Somogyi, G.},
  \bibinfo{author}{Szor, Z.}, \bibinfo{author}{{Tr\'ocs\'anyi}, Z.},
  \bibinfo{author}{{Tulip\'ant}, Z.}, \bibinfo{year}{2016}.
\newblock \bibinfo{title}{{Jet production in the CoLoRFulNNLO method: event
  shapes in electron-positron collisions}}.
\newblock \bibinfo{journal}{Phys. Rev. D} \bibinfo{volume}{94},
  \bibinfo{pages}{074019}.
\newblock \DOIprefix\doi{10.1103/PhysRevD.94.074019},
  \href{http://arxiv.org/abs/1606.03453}{{\tt arXiv:1606.03453}}.
%Type = Article
\bibitem[{d'Enterria and Monni()}]{2921876}
\bibinfo{author}{d'Enterria, D.}, \bibinfo{author}{Monni, P.F.}
  (\bibinfo{collaboration}{FCC}), .
\newblock \bibinfo{title}{{QCD physics at FCC}}
  \DOIprefix\doi{10.17181/nfcpy-vns54}.
%Type = Article
\bibitem[{Devoto et~al.(2024)Devoto, Melnikov, R\"ontsch, Signorile-Signorile
  and Tagliabue}]{Devoto:2023rpv}
\bibinfo{author}{Devoto, F.}, \bibinfo{author}{Melnikov, K.},
  \bibinfo{author}{R\"ontsch, R.}, \bibinfo{author}{Signorile-Signorile, C.},
  \bibinfo{author}{Tagliabue, D.M.}, \bibinfo{year}{2024}.
\newblock \bibinfo{title}{{A fresh look at the nested soft-collinear
  subtraction scheme: NNLO QCD corrections to N-gluon final states in $
  q\overline{q} $ annihilation}}.
\newblock \bibinfo{journal}{JHEP} \bibinfo{volume}{02}, \bibinfo{pages}{016}.
\newblock \DOIprefix\doi{10.1007/JHEP02(2024)016},
  \href{http://arxiv.org/abs/2310.17598}{{\tt arXiv:2310.17598}}.
%Type = Article
\bibitem[{Devoto et~al.(2025)Devoto, Melnikov, R\"ontsch, Signorile-Signorile,
  Tagliabue and Tresoldi}]{Devoto:2025kin}
\bibinfo{author}{Devoto, F.}, \bibinfo{author}{Melnikov, K.},
  \bibinfo{author}{R\"ontsch, R.}, \bibinfo{author}{Signorile-Signorile, C.},
  \bibinfo{author}{Tagliabue, D.M.}, \bibinfo{author}{Tresoldi, M.},
  \bibinfo{year}{2025}.
\newblock \bibinfo{title}{{Towards a general subtraction formula for NNLO QCD
  corrections to processes at hadron colliders: final states with quarks and
  gluons}} \href{http://arxiv.org/abs/2503.15251}{{\tt arXiv:2503.15251}}.
%Type = Article
\bibitem[{Dixon and Signer(1997)}]{Dixon:1997th}
\bibinfo{author}{Dixon, L.J.}, \bibinfo{author}{Signer, A.},
  \bibinfo{year}{1997}.
\newblock \bibinfo{title}{{Complete $\mathcal{O}(\alpha_s^3)$ results for $e^+
  e^- \to (\gamma, Z) \to$ four jets}}.
\newblock \bibinfo{journal}{Phys. Rev. D} \bibinfo{volume}{56},
  \bibinfo{pages}{4031--4038}.
\newblock \DOIprefix\doi{10.1103/PhysRevD.56.4031},
  \href{http://arxiv.org/abs/hep-ph/9706285}{{\tt arXiv:hep-ph/9706285}}.
%Type = Article
\bibitem[{Dreyer and Karlberg(2016)}]{Dreyer:2016oyx}
\bibinfo{author}{Dreyer, F.A.}, \bibinfo{author}{Karlberg, A.},
  \bibinfo{year}{2016}.
\newblock \bibinfo{title}{{Vector-Boson Fusion Higgs Production at Three Loops
  in QCD}}.
\newblock \bibinfo{journal}{Phys. Rev. Lett.} \bibinfo{volume}{117},
  \bibinfo{pages}{072001}.
\newblock \DOIprefix\doi{10.1103/PhysRevLett.117.072001},
  \href{http://arxiv.org/abs/1606.00840}{{\tt arXiv:1606.00840}}.
%Type = Article
\bibitem[{Dreyer and Karlberg(2018)}]{Dreyer:2018qbw}
\bibinfo{author}{Dreyer, F.A.}, \bibinfo{author}{Karlberg, A.},
  \bibinfo{year}{2018}.
\newblock \bibinfo{title}{{Vector-Boson Fusion Higgs Pair Production at
  N$^3$LO}}.
\newblock \bibinfo{journal}{Phys. Rev. D} \bibinfo{volume}{98},
  \bibinfo{pages}{114016}.
\newblock \DOIprefix\doi{10.1103/PhysRevD.98.114016},
  \href{http://arxiv.org/abs/1811.07906}{{\tt arXiv:1811.07906}}.
%Type = Article
\bibitem[{Duhr and Dulat(2019)}]{Duhr:2019tlz}
\bibinfo{author}{Duhr, C.}, \bibinfo{author}{Dulat, F.}, \bibinfo{year}{2019}.
\newblock \bibinfo{title}{{PolyLogTools \textemdash{} polylogs for the
  masses}}.
\newblock \bibinfo{journal}{JHEP} \bibinfo{volume}{08}, \bibinfo{pages}{135}.
\newblock \DOIprefix\doi{10.1007/JHEP08(2019)135},
  \href{http://arxiv.org/abs/1904.07279}{{\tt arXiv:1904.07279}}.
%Type = Article
\bibitem[{Duhr et~al.(2020a)Duhr, Dulat and Mistlberger}]{Duhr:2020sdp}
\bibinfo{author}{Duhr, C.}, \bibinfo{author}{Dulat, F.},
  \bibinfo{author}{Mistlberger, B.}, \bibinfo{year}{2020}a.
\newblock \bibinfo{title}{{Charged current Drell-Yan production at N$^{3}$LO}}.
\newblock \bibinfo{journal}{JHEP} \bibinfo{volume}{11}, \bibinfo{pages}{143}.
\newblock \DOIprefix\doi{10.1007/JHEP11(2020)143},
  \href{http://arxiv.org/abs/2007.13313}{{\tt arXiv:2007.13313}}.
%Type = Article
\bibitem[{Duhr et~al.(2020b)Duhr, Dulat and Mistlberger}]{Duhr:2020seh}
\bibinfo{author}{Duhr, C.}, \bibinfo{author}{Dulat, F.},
  \bibinfo{author}{Mistlberger, B.}, \bibinfo{year}{2020}b.
\newblock \bibinfo{title}{{Drell-Yan Cross Section to Third Order in the Strong
  Coupling Constant}}.
\newblock \bibinfo{journal}{Phys. Rev. Lett.} \bibinfo{volume}{125},
  \bibinfo{pages}{172001}.
\newblock \DOIprefix\doi{10.1103/PhysRevLett.125.172001},
  \href{http://arxiv.org/abs/2001.07717}{{\tt arXiv:2001.07717}}.
%Type = Article
\bibitem[{Duhr and Mistlberger(2022)}]{Duhr:2021vwj}
\bibinfo{author}{Duhr, C.}, \bibinfo{author}{Mistlberger, B.},
  \bibinfo{year}{2022}.
\newblock \bibinfo{title}{{Lepton-pair production at hadron colliders at
  N$^{3}$LO in QCD}}.
\newblock \bibinfo{journal}{JHEP} \bibinfo{volume}{03}, \bibinfo{pages}{116}.
\newblock \DOIprefix\doi{10.1007/JHEP03(2022)116},
  \href{http://arxiv.org/abs/2111.10379}{{\tt arXiv:2111.10379}}.
%Type = Article
\bibitem[{Falck et~al.(1989)Falck, Graudenz and Kramer}]{Falck:1988gc}
\bibinfo{author}{Falck, N.K.}, \bibinfo{author}{Graudenz, D.},
  \bibinfo{author}{Kramer, G.}, \bibinfo{year}{1989}.
\newblock \bibinfo{title}{{Five Jet Production in $e^+ e^-$ Annihilation}}.
\newblock \bibinfo{journal}{Phys. Lett. B} \bibinfo{volume}{220},
  \bibinfo{pages}{299--302}.
\newblock \DOIprefix\doi{10.1016/0370-2693(89)90056-7}.
%Type = Article
\bibitem[{Fontana et~al.(2024)Fontana, Gehrmann and
  Sch\"onwald}]{Fontana:2024olm}
\bibinfo{author}{Fontana, G.}, \bibinfo{author}{Gehrmann, T.},
  \bibinfo{author}{Sch\"onwald, K.}, \bibinfo{year}{2024}.
\newblock \bibinfo{title}{{Analytic auxiliary mass flow to compute master
  integrals in singular kinematics}}.
\newblock \bibinfo{journal}{JHEP} \bibinfo{volume}{03}, \bibinfo{pages}{159}.
\newblock \DOIprefix\doi{10.1007/JHEP03(2024)159},
  \href{http://arxiv.org/abs/2401.08226}{{\tt arXiv:2401.08226}}.
%Type = Article
\bibitem[{Fox et~al.(2025a)Fox, Gehrmann-De~Ridder, Gehrmann, Glover, Marcoli
  and Preuss}]{Fox:2025cuz}
\bibinfo{author}{Fox, E.}, \bibinfo{author}{Gehrmann-De~Ridder, A.},
  \bibinfo{author}{Gehrmann, T.}, \bibinfo{author}{Glover, N.},
  \bibinfo{author}{Marcoli, M.}, \bibinfo{author}{Preuss, C.T.},
  \bibinfo{year}{2025}a.
\newblock \bibinfo{title}{{Jet rates in Higgs boson decay at third order in
  QCD}} \href{http://arxiv.org/abs/2502.17333}{{\tt arXiv:2502.17333}}.
%Type = Article
\bibitem[{Fox and Glover(2023)}]{Fox:2023bma}
\bibinfo{author}{Fox, E.}, \bibinfo{author}{Glover, N.}, \bibinfo{year}{2023}.
\newblock \bibinfo{title}{{Initial-final and initial-initial antenna functions
  for real radiation at next-to-leading order}}.
\newblock \bibinfo{journal}{JHEP} \bibinfo{volume}{12}, \bibinfo{pages}{171}.
\newblock \DOIprefix\doi{10.1007/JHEP12(2023)171},
  \href{http://arxiv.org/abs/2308.10829}{{\tt arXiv:2308.10829}}.
%Type = Article
\bibitem[{Fox et~al.(2025b)Fox, Glover and Marcoli}]{Fox:2024bfp}
\bibinfo{author}{Fox, E.}, \bibinfo{author}{Glover, N.},
  \bibinfo{author}{Marcoli, M.}, \bibinfo{year}{2025}b.
\newblock \bibinfo{title}{{Generalised antenna functions for higher-order
  calculations}}.
\newblock \bibinfo{journal}{JHEP} \bibinfo{volume}{12}, \bibinfo{pages}{225}.
\newblock \DOIprefix\doi{10.1007/JHEP12(2024)225},
  \href{http://arxiv.org/abs/2410.12904}{{\tt arXiv:2410.12904}}.
%Type = Article
\bibitem[{Frederix et~al.(2010)Frederix, Frixione, Melnikov and
  Zanderighi}]{Frederix:2010ne}
\bibinfo{author}{Frederix, R.}, \bibinfo{author}{Frixione, S.},
  \bibinfo{author}{Melnikov, K.}, \bibinfo{author}{Zanderighi, G.},
  \bibinfo{year}{2010}.
\newblock \bibinfo{title}{{NLO QCD corrections to five-jet production at LEP
  and the extraction of $\alpha_s(M_Z)$}}.
\newblock \bibinfo{journal}{JHEP} \bibinfo{volume}{11}, \bibinfo{pages}{050}.
\newblock \DOIprefix\doi{10.1007/JHEP11(2010)050},
  \href{http://arxiv.org/abs/1008.5313}{{\tt arXiv:1008.5313}}.
%Type = Article
\bibitem[{Frixione et~al.(1996)Frixione, Kunszt and Signer}]{Frixione:1995ms}
\bibinfo{author}{Frixione, S.}, \bibinfo{author}{Kunszt, Z.},
  \bibinfo{author}{Signer, A.}, \bibinfo{year}{1996}.
\newblock \bibinfo{title}{{Three jet cross-sections to next-to-leading order}}.
\newblock \bibinfo{journal}{Nucl. Phys. B} \bibinfo{volume}{467},
  \bibinfo{pages}{399--442}.
\newblock \DOIprefix\doi{10.1016/0550-3213(96)00110-1},
  \href{http://arxiv.org/abs/hep-ph/9512328}{{\tt arXiv:hep-ph/9512328}}.
%Type = Article
\bibitem[{Garland et~al.(2002a)Garland, Gehrmann, Glover, Koukoutsakis and
  Remiddi}]{Garland:2001tf}
\bibinfo{author}{Garland, L.W.}, \bibinfo{author}{Gehrmann, T.},
  \bibinfo{author}{Glover, E.W.N.}, \bibinfo{author}{Koukoutsakis, A.},
  \bibinfo{author}{Remiddi, E.}, \bibinfo{year}{2002}a.
\newblock \bibinfo{title}{{The Two loop QCD matrix element for $e^+ e^- \to$ 3
  jets}}.
\newblock \bibinfo{journal}{Nucl. Phys. B} \bibinfo{volume}{627},
  \bibinfo{pages}{107--188}.
\newblock \DOIprefix\doi{10.1016/S0550-3213(02)00057-3},
  \href{http://arxiv.org/abs/hep-ph/0112081}{{\tt arXiv:hep-ph/0112081}}.
%Type = Article
\bibitem[{Garland et~al.(2002b)Garland, Gehrmann, Glover, Koukoutsakis and
  Remiddi}]{Garland:2002ak}
\bibinfo{author}{Garland, L.W.}, \bibinfo{author}{Gehrmann, T.},
  \bibinfo{author}{Glover, E.W.N.}, \bibinfo{author}{Koukoutsakis, A.},
  \bibinfo{author}{Remiddi, E.}, \bibinfo{year}{2002}b.
\newblock \bibinfo{title}{{Two loop QCD helicity amplitudes for $e^+ e^- \to$
  three jets}}.
\newblock \bibinfo{journal}{Nucl. Phys. B} \bibinfo{volume}{642},
  \bibinfo{pages}{227--262}.
\newblock \DOIprefix\doi{10.1016/S0550-3213(02)00627-2},
  \href{http://arxiv.org/abs/hep-ph/0206067}{{\tt arXiv:hep-ph/0206067}}.
%Type = Article
\bibitem[{Gaunt et~al.(2015)Gaunt, Stahlhofen, Tackmann and
  Walsh}]{Gaunt:2015pea}
\bibinfo{author}{Gaunt, J.}, \bibinfo{author}{Stahlhofen, M.},
  \bibinfo{author}{Tackmann, F.J.}, \bibinfo{author}{Walsh, J.R.},
  \bibinfo{year}{2015}.
\newblock \bibinfo{title}{{N-jettiness Subtractions for NNLO QCD
  Calculations}}.
\newblock \bibinfo{journal}{JHEP} \bibinfo{volume}{09}, \bibinfo{pages}{058}.
\newblock \DOIprefix\doi{10.1007/JHEP09(2015)058},
  \href{http://arxiv.org/abs/1505.04794}{{\tt arXiv:1505.04794}}.
%Type = Article
\bibitem[{Gehrmann et~al.(2010)Gehrmann, Glover, Huber, Ikizlerli and
  Studerus}]{Gehrmann:2010ue}
\bibinfo{author}{Gehrmann, T.}, \bibinfo{author}{Glover, E.W.N.},
  \bibinfo{author}{Huber, T.}, \bibinfo{author}{Ikizlerli, N.},
  \bibinfo{author}{Studerus, C.}, \bibinfo{year}{2010}.
\newblock \bibinfo{title}{{Calculation of the quark and gluon form factors to
  three loops in QCD}}.
\newblock \bibinfo{journal}{JHEP} \bibinfo{volume}{06}, \bibinfo{pages}{094}.
\newblock \DOIprefix\doi{10.1007/JHEP06(2010)094},
  \href{http://arxiv.org/abs/1004.3653}{{\tt arXiv:1004.3653}}.
%Type = Article
\bibitem[{Gehrmann et~al.(2017)Gehrmann, Glover, Huss, Niehues and
  Zhang}]{Gehrmann:2017xfb}
\bibinfo{author}{Gehrmann, T.}, \bibinfo{author}{Glover, E.W.N.},
  \bibinfo{author}{Huss, A.}, \bibinfo{author}{Niehues, J.},
  \bibinfo{author}{Zhang, H.}, \bibinfo{year}{2017}.
\newblock \bibinfo{title}{{NNLO QCD corrections to event orientation in $e^+
  e^-$ annihilation}}.
\newblock \bibinfo{journal}{Phys. Lett. B} \bibinfo{volume}{775},
  \bibinfo{pages}{185--189}.
\newblock \DOIprefix\doi{10.1016/j.physletb.2017.10.069},
  \href{http://arxiv.org/abs/1709.01097}{{\tt arXiv:1709.01097}}.
%Type = Article
\bibitem[{Gehrmann et~al.(2024)Gehrmann, Glover and Marcoli}]{Gehrmann:2023dxm}
\bibinfo{author}{Gehrmann, T.}, \bibinfo{author}{Glover, E.W.N.},
  \bibinfo{author}{Marcoli, M.}, \bibinfo{year}{2024}.
\newblock \bibinfo{title}{{The colourful antenna subtraction method}}.
\newblock \bibinfo{journal}{JHEP} \bibinfo{volume}{03}, \bibinfo{pages}{114}.
\newblock \DOIprefix\doi{10.1007/JHEP03(2024)114},
  \href{http://arxiv.org/abs/2310.19757}{{\tt arXiv:2310.19757}}.
%Type = Article
\bibitem[{Gehrmann and Remiddi(2001)}]{Gehrmann:2000zt}
\bibinfo{author}{Gehrmann, T.}, \bibinfo{author}{Remiddi, E.},
  \bibinfo{year}{2001}.
\newblock \bibinfo{title}{{Two loop master integrals for $\gamma^* \to$ 3 jets:
  The Planar topologies}}.
\newblock \bibinfo{journal}{Nucl. Phys. B} \bibinfo{volume}{601},
  \bibinfo{pages}{248--286}.
\newblock \DOIprefix\doi{10.1016/S0550-3213(01)00057-8},
  \href{http://arxiv.org/abs/hep-ph/0008287}{{\tt arXiv:hep-ph/0008287}}.
%Type = Article
\bibitem[{Gehrmann and Remiddi(2002)}]{Gehrmann:2001jv}
\bibinfo{author}{Gehrmann, T.}, \bibinfo{author}{Remiddi, E.},
  \bibinfo{year}{2002}.
\newblock \bibinfo{title}{{Numerical evaluation of two-dimensional harmonic
  polylogarithms}}.
\newblock \bibinfo{journal}{Comput. Phys. Commun.} \bibinfo{volume}{144},
  \bibinfo{pages}{200--223}.
\newblock \DOIprefix\doi{10.1016/S0010-4655(02)00139-X},
  \href{http://arxiv.org/abs/hep-ph/0111255}{{\tt arXiv:hep-ph/0111255}}.
%Type = Article
\bibitem[{Gehrmann-De~Ridder et~al.(2004)Gehrmann-De~Ridder, Gehrmann and
  Glover}]{Gehrmann-DeRidder:2004ttg}
\bibinfo{author}{Gehrmann-De~Ridder, A.}, \bibinfo{author}{Gehrmann, T.},
  \bibinfo{author}{Glover, E.W.N.}, \bibinfo{year}{2004}.
\newblock \bibinfo{title}{{Infrared structure of $e^+e^-\to$ 2 jets at NNLO}}.
\newblock \bibinfo{journal}{Nucl. Phys. B} \bibinfo{volume}{691},
  \bibinfo{pages}{195--222}.
\newblock \DOIprefix\doi{10.1016/j.nuclphysb.2004.05.017},
  \href{http://arxiv.org/abs/hep-ph/0403057}{{\tt arXiv:hep-ph/0403057}}.
%Type = Article
\bibitem[{Gehrmann-De~Ridder et~al.(2005a)Gehrmann-De~Ridder, Gehrmann and
  Glover}]{Gehrmann-DeRidder:2005btv}
\bibinfo{author}{Gehrmann-De~Ridder, A.}, \bibinfo{author}{Gehrmann, T.},
  \bibinfo{author}{Glover, E.W.N.}, \bibinfo{year}{2005}a.
\newblock \bibinfo{title}{{Antenna subtraction at NNLO}}.
\newblock \bibinfo{journal}{JHEP} \bibinfo{volume}{09}, \bibinfo{pages}{056}.
\newblock \DOIprefix\doi{10.1088/1126-6708/2005/09/056},
  \href{http://arxiv.org/abs/hep-ph/0505111}{{\tt arXiv:hep-ph/0505111}}.
%Type = Article
\bibitem[{Gehrmann-De~Ridder et~al.(2005b)Gehrmann-De~Ridder, Gehrmann and
  Glover}]{Gehrmann-DeRidder:2005alt}
\bibinfo{author}{Gehrmann-De~Ridder, A.}, \bibinfo{author}{Gehrmann, T.},
  \bibinfo{author}{Glover, E.W.N.}, \bibinfo{year}{2005}b.
\newblock \bibinfo{title}{{Gluon-gluon antenna functions from Higgs boson
  decay}}.
\newblock \bibinfo{journal}{Phys. Lett. B} \bibinfo{volume}{612},
  \bibinfo{pages}{49--60}.
\newblock \DOIprefix\doi{10.1016/j.physletb.2005.03.003},
  \href{http://arxiv.org/abs/hep-ph/0502110}{{\tt arXiv:hep-ph/0502110}}.
%Type = Article
\bibitem[{Gehrmann-De~Ridder et~al.(2005c)Gehrmann-De~Ridder, Gehrmann and
  Glover}]{Gehrmann-DeRidder:2005svg}
\bibinfo{author}{Gehrmann-De~Ridder, A.}, \bibinfo{author}{Gehrmann, T.},
  \bibinfo{author}{Glover, E.W.N.}, \bibinfo{year}{2005}c.
\newblock \bibinfo{title}{{Quark-gluon antenna functions from neutralino
  decay}}.
\newblock \bibinfo{journal}{Phys. Lett. B} \bibinfo{volume}{612},
  \bibinfo{pages}{36--48}.
\newblock \DOIprefix\doi{10.1016/j.physletb.2005.02.039},
  \href{http://arxiv.org/abs/hep-ph/0501291}{{\tt arXiv:hep-ph/0501291}}.
%Type = Article
\bibitem[{Gehrmann-De~Ridder et~al.(2007)Gehrmann-De~Ridder, Gehrmann, Glover
  and Heinrich}]{Gehrmann-DeRidder:2007foh}
\bibinfo{author}{Gehrmann-De~Ridder, A.}, \bibinfo{author}{Gehrmann, T.},
  \bibinfo{author}{Glover, E.W.N.}, \bibinfo{author}{Heinrich, G.},
  \bibinfo{year}{2007}.
\newblock \bibinfo{title}{{Infrared structure of $e^+ e^- \to$ 3 jets at
  NNLO}}.
\newblock \bibinfo{journal}{JHEP} \bibinfo{volume}{11}, \bibinfo{pages}{058}.
\newblock \DOIprefix\doi{10.1088/1126-6708/2007/11/058},
  \href{http://arxiv.org/abs/0710.0346}{{\tt arXiv:0710.0346}}.
%Type = Article
\bibitem[{Gehrmann-De~Ridder et~al.(2008)Gehrmann-De~Ridder, Gehrmann, Glover
  and Heinrich}]{Gehrmann-DeRidder:2008qsl}
\bibinfo{author}{Gehrmann-De~Ridder, A.}, \bibinfo{author}{Gehrmann, T.},
  \bibinfo{author}{Glover, E.W.N.}, \bibinfo{author}{Heinrich, G.},
  \bibinfo{year}{2008}.
\newblock \bibinfo{title}{{Jet rates in electron-positron annihilation at
  $\mathcal{O}(\alpha_s^3)$ in QCD}}.
\newblock \bibinfo{journal}{Phys. Rev. Lett.} \bibinfo{volume}{100},
  \bibinfo{pages}{172001}.
\newblock \DOIprefix\doi{10.1103/PhysRevLett.100.172001},
  \href{http://arxiv.org/abs/0802.0813}{{\tt arXiv:0802.0813}}.
%Type = Article
\bibitem[{Gehrmann-De~Ridder et~al.(2014)Gehrmann-De~Ridder, Gehrmann, Glover
  and Heinrich}]{Gehrmann-DeRidder:2014hxk}
\bibinfo{author}{Gehrmann-De~Ridder, A.}, \bibinfo{author}{Gehrmann, T.},
  \bibinfo{author}{Glover, E.W.N.}, \bibinfo{author}{Heinrich, G.},
  \bibinfo{year}{2014}.
\newblock \bibinfo{title}{{EERAD3: Event shapes and jet rates in
  electron-positron annihilation at order $\alpha_s^3$}}.
\newblock \bibinfo{journal}{Comput. Phys. Commun.} \bibinfo{volume}{185},
  \bibinfo{pages}{3331}.
\newblock \DOIprefix\doi{10.1016/j.cpc.2014.07.024},
  \href{http://arxiv.org/abs/1402.4140}{{\tt arXiv:1402.4140}}.
%Type = Article
\bibitem[{Glover and Miller(1997)}]{Glover:1996eh}
\bibinfo{author}{Glover, E.W.N.}, \bibinfo{author}{Miller, D.J.},
  \bibinfo{year}{1997}.
\newblock \bibinfo{title}{{The One loop QCD corrections for $gamma^* \to
  Q\bar{Q}q\bar{q}$}}.
\newblock \bibinfo{journal}{Phys. Lett. B} \bibinfo{volume}{396},
  \bibinfo{pages}{257--263}.
\newblock \DOIprefix\doi{10.1016/S0370-2693(97)00113-5},
  \href{http://arxiv.org/abs/hep-ph/9609474}{{\tt arXiv:hep-ph/9609474}}.
%Type = Article
\bibitem[{Glover and Pires(2010)}]{NigelGlover:2010kwr}
\bibinfo{author}{Glover, E.W.N.}, \bibinfo{author}{Pires, J.},
  \bibinfo{year}{2010}.
\newblock \bibinfo{title}{{Antenna subtraction for gluon scattering at NNLO}}.
\newblock \bibinfo{journal}{JHEP} \bibinfo{volume}{06}, \bibinfo{pages}{096}.
\newblock \DOIprefix\doi{10.1007/JHEP06(2010)096},
  \href{http://arxiv.org/abs/1003.2824}{{\tt arXiv:1003.2824}}.
%Type = Article
\bibitem[{Hagiwara and Zeppenfeld(1989)}]{Hagiwara:1988pp}
\bibinfo{author}{Hagiwara, K.}, \bibinfo{author}{Zeppenfeld, D.},
  \bibinfo{year}{1989}.
\newblock \bibinfo{title}{{Amplitudes for Multiparton Processes Involving a
  Current at $e^+ e^-$, $e^{\pm} p$, and Hadron Colliders}}.
\newblock \bibinfo{journal}{Nucl. Phys. B} \bibinfo{volume}{313},
  \bibinfo{pages}{560--594}.
\newblock \DOIprefix\doi{10.1016/0550-3213(89)90397-0}.
%Type = Article
\bibitem[{He et~al.(2025)He, Xing, Yang and Zhu}]{He:2025hin}
\bibinfo{author}{He, C.Q.}, \bibinfo{author}{Xing, H.}, \bibinfo{author}{Yang,
  T.Z.}, \bibinfo{author}{Zhu, H.X.}, \bibinfo{year}{2025}.
\newblock \bibinfo{title}{{Single-inclusive hadron production in
  electron-positron annihilation at next-to-next-to-next-to-leading order in
  QCD}} \href{http://arxiv.org/abs/2503.20441}{{\tt arXiv:2503.20441}}.
%Type = Article
\bibitem[{Herzog(2018)}]{Herzog:2018ily}
\bibinfo{author}{Herzog, F.}, \bibinfo{year}{2018}.
\newblock \bibinfo{title}{{Geometric IR subtraction for final state real
  radiation}}.
\newblock \bibinfo{journal}{JHEP} \bibinfo{volume}{08}, \bibinfo{pages}{006}.
\newblock \DOIprefix\doi{10.1007/JHEP08(2018)006},
  \href{http://arxiv.org/abs/1804.07949}{{\tt arXiv:1804.07949}}.
%Type = Article
\bibitem[{Huss et~al.(2023)Huss, Huston, Jones and Pellen}]{Huss:2022ful}
\bibinfo{author}{Huss, A.}, \bibinfo{author}{Huston, J.},
  \bibinfo{author}{Jones, S.}, \bibinfo{author}{Pellen, M.},
  \bibinfo{year}{2023}.
\newblock \bibinfo{title}{{Les Houches 2021\textemdash{}physics at TeV
  colliders: report on the standard model precision wishlist}}.
\newblock \bibinfo{journal}{J. Phys. G} \bibinfo{volume}{50},
  \bibinfo{pages}{043001}.
\newblock \DOIprefix\doi{10.1088/1361-6471/acbaec},
  \href{http://arxiv.org/abs/2207.02122}{{\tt arXiv:2207.02122}}.
%Type = Article
\bibitem[{Huss et~al.(2025)}]{NNLOJET:2025rno}
\bibinfo{author}{Huss, A.}, et~al. (\bibinfo{collaboration}{NNLOJET}),
  \bibinfo{year}{2025}.
\newblock \bibinfo{title}{{NNLOJET: a parton-level event generator for jet
  cross sections at NNLO QCD accuracy}}
  \href{http://arxiv.org/abs/2503.22804}{{\tt arXiv:2503.22804}}.
%Type = Article
\bibitem[{Jakub\v{c}\'\i{}k et~al.(2023)Jakub\v{c}\'\i{}k, Marcoli and
  Stagnitto}]{Jakubcik:2022zdi}
\bibinfo{author}{Jakub\v{c}\'\i{}k, P.}, \bibinfo{author}{Marcoli, M.},
  \bibinfo{author}{Stagnitto, G.}, \bibinfo{year}{2023}.
\newblock \bibinfo{title}{{The parton-level structure of e$^{+}$e$^{-}$ to 2
  jets at N$^{3}$LO}}.
\newblock \bibinfo{journal}{JHEP} \bibinfo{volume}{01}, \bibinfo{pages}{168}.
\newblock \DOIprefix\doi{10.1007/JHEP01(2023)168},
  \href{http://arxiv.org/abs/2211.08446}{{\tt arXiv:2211.08446}}.
%Type = Article
\bibitem[{Kinoshita(1962)}]{Kinoshita:1962ur}
\bibinfo{author}{Kinoshita, T.}, \bibinfo{year}{1962}.
\newblock \bibinfo{title}{{Mass singularities of Feynman amplitudes}}.
\newblock \bibinfo{journal}{J. Math. Phys.} \bibinfo{volume}{3},
  \bibinfo{pages}{650--677}.
\newblock \DOIprefix\doi{10.1063/1.1724268}.
%Type = Article
\bibitem[{Kosower(1998)}]{Kosower:1997zr}
\bibinfo{author}{Kosower, D.A.}, \bibinfo{year}{1998}.
\newblock \bibinfo{title}{{Antenna factorization of gauge theory amplitudes}}.
\newblock \bibinfo{journal}{Phys. Rev. D} \bibinfo{volume}{57},
  \bibinfo{pages}{5410--5416}.
\newblock \DOIprefix\doi{10.1103/PhysRevD.57.5410},
  \href{http://arxiv.org/abs/hep-ph/9710213}{{\tt arXiv:hep-ph/9710213}}.
%Type = Article
\bibitem[{Kosower(2003)}]{Kosower:2002su}
\bibinfo{author}{Kosower, D.A.}, \bibinfo{year}{2003}.
\newblock \bibinfo{title}{{Multiple singular emission in gauge theories}}.
\newblock \bibinfo{journal}{Phys. Rev. D} \bibinfo{volume}{67},
  \bibinfo{pages}{116003}.
\newblock \DOIprefix\doi{10.1103/PhysRevD.67.116003},
  \href{http://arxiv.org/abs/hep-ph/0212097}{{\tt arXiv:hep-ph/0212097}}.
%Type = Article
\bibitem[{Lee et~al.(2010)Lee, Smirnov and Smirnov}]{Lee:2010cga}
\bibinfo{author}{Lee, R.N.}, \bibinfo{author}{Smirnov, A.V.},
  \bibinfo{author}{Smirnov, V.A.}, \bibinfo{year}{2010}.
\newblock \bibinfo{title}{{Analytic Results for Massless Three-Loop Form
  Factors}}.
\newblock \bibinfo{journal}{JHEP} \bibinfo{volume}{04}, \bibinfo{pages}{020}.
\newblock \DOIprefix\doi{10.1007/JHEP04(2010)020},
  \href{http://arxiv.org/abs/1001.2887}{{\tt arXiv:1001.2887}}.
%Type = Article
\bibitem[{Lee and Nauenberg(1964)}]{Lee:1964is}
\bibinfo{author}{Lee, T.D.}, \bibinfo{author}{Nauenberg, M.},
  \bibinfo{year}{1964}.
\newblock \bibinfo{title}{{Degenerate Systems and Mass Singularities}}.
\newblock \bibinfo{journal}{Phys. Rev.} \bibinfo{volume}{133},
  \bibinfo{pages}{B1549--B1562}.
\newblock \DOIprefix\doi{10.1103/PhysRev.133.B1549}.
%Type = Article
\bibitem[{Magnea et~al.(2018)Magnea, Maina, Pelliccioli, Signorile-Signorile,
  Torrielli and Uccirati}]{Magnea:2018hab}
\bibinfo{author}{Magnea, L.}, \bibinfo{author}{Maina, E.},
  \bibinfo{author}{Pelliccioli, G.}, \bibinfo{author}{Signorile-Signorile, C.},
  \bibinfo{author}{Torrielli, P.}, \bibinfo{author}{Uccirati, S.},
  \bibinfo{year}{2018}.
\newblock \bibinfo{title}{{Local analytic sector subtraction at NNLO}}.
\newblock \bibinfo{journal}{JHEP} \bibinfo{volume}{12}, \bibinfo{pages}{107}.
\newblock \DOIprefix\doi{10.1007/JHEP12(2018)107},
  \href{http://arxiv.org/abs/1806.09570}{{\tt arXiv:1806.09570}}.
  \bibinfo{note}{[Erratum: JHEP 06, 013 (2019)]}.
%Type = Article
\bibitem[{Mistlberger(2018)}]{Mistlberger:2018etf}
\bibinfo{author}{Mistlberger, B.}, \bibinfo{year}{2018}.
\newblock \bibinfo{title}{{Higgs boson production at hadron colliders at
  N$^{3}$LO in QCD}}.
\newblock \bibinfo{journal}{JHEP} \bibinfo{volume}{05}, \bibinfo{pages}{028}.
\newblock \DOIprefix\doi{10.1007/JHEP05(2018)028},
  \href{http://arxiv.org/abs/1802.00833}{{\tt arXiv:1802.00833}}.
%Type = Article
\bibitem[{Moch et~al.(2002)Moch, Uwer and Weinzierl}]{Moch:2001zr}
\bibinfo{author}{Moch, S.}, \bibinfo{author}{Uwer, P.},
  \bibinfo{author}{Weinzierl, S.}, \bibinfo{year}{2002}.
\newblock \bibinfo{title}{{Nested sums, expansion of transcendental functions
  and multiscale multiloop integrals}}.
\newblock \bibinfo{journal}{J. Math. Phys.} \bibinfo{volume}{43},
  \bibinfo{pages}{3363--3386}.
\newblock \DOIprefix\doi{10.1063/1.1471366},
  \href{http://arxiv.org/abs/hep-ph/0110083}{{\tt arXiv:hep-ph/0110083}}.
%Type = Article
\bibitem[{Mondini et~al.(2019)Mondini, Schiavi and Williams}]{Mondini:2019gid}
\bibinfo{author}{Mondini, R.}, \bibinfo{author}{Schiavi, M.},
  \bibinfo{author}{Williams, C.}, \bibinfo{year}{2019}.
\newblock \bibinfo{title}{{N$^{3}$LO predictions for the decay of the Higgs
  boson to bottom quarks}}.
\newblock \bibinfo{journal}{JHEP} \bibinfo{volume}{06}, \bibinfo{pages}{079}.
\newblock \DOIprefix\doi{10.1007/JHEP06(2019)079},
  \href{http://arxiv.org/abs/1904.08960}{{\tt arXiv:1904.08960}}.
%Type = Article
\bibitem[{Neumann and Campbell(2023)}]{Neumann:2022lft}
\bibinfo{author}{Neumann, T.}, \bibinfo{author}{Campbell, J.},
  \bibinfo{year}{2023}.
\newblock \bibinfo{title}{{Fiducial Drell-Yan production at the LHC improved by
  transverse-momentum resummation at N4LLp+N3LO}}.
\newblock \bibinfo{journal}{Phys. Rev. D} \bibinfo{volume}{107},
  \bibinfo{pages}{L011506}.
\newblock \DOIprefix\doi{10.1103/PhysRevD.107.L011506},
  \href{http://arxiv.org/abs/2207.07056}{{\tt arXiv:2207.07056}}.
%Type = Article
\bibitem[{Stirling(1991)}]{Stirling:1991ds}
\bibinfo{author}{Stirling, W.J.}, \bibinfo{year}{1991}.
\newblock \bibinfo{title}{{Hard QCD working group: Theory summary}}.
\newblock \bibinfo{journal}{J. Phys. G} \bibinfo{volume}{17},
  \bibinfo{pages}{1567--1574}.
\newblock \DOIprefix\doi{10.1088/0954-3899/17/10/014}.
%Type = Article
\bibitem[{Torres~Bobadilla et~al.(2021)}]{TorresBobadilla:2020ekr}
\bibinfo{author}{Torres~Bobadilla, W.J.}, et~al., \bibinfo{year}{2021}.
\newblock \bibinfo{title}{{May the four be with you: Novel IR-subtraction
  methods to tackle NNLO calculations}}.
\newblock \bibinfo{journal}{Eur. Phys. J. C} \bibinfo{volume}{81},
  \bibinfo{pages}{250}.
\newblock \DOIprefix\doi{10.1140/epjc/s10052-021-08996-y},
  \href{http://arxiv.org/abs/2012.02567}{{\tt arXiv:2012.02567}}.
%Type = Inproceedings
\bibitem[{Verbytskyi et~al.(2025)Verbytskyi, d'Enterria, Monni and
  Skands}]{Verbytskyi:2025sod}
\bibinfo{author}{Verbytskyi, A.}, \bibinfo{author}{d'Enterria, D.},
  \bibinfo{author}{Monni, P.F.}, \bibinfo{author}{Skands, P.},
  \bibinfo{year}{2025}.
\newblock \bibinfo{title}{{Physics case for low-$\sqrt{s}$ QCD studies at
  FCC-ee}}.
\newblock \href{http://arxiv.org/abs/2503.23855}{{\tt arXiv:2503.23855}}.
%Type = Article
\bibitem[{Weinzierl(2006)}]{Weinzierl:2006ij}
\bibinfo{author}{Weinzierl, S.}, \bibinfo{year}{2006}.
\newblock \bibinfo{title}{{NNLO corrections to 2-jet observables in
  electron-positron annihilation}}.
\newblock \bibinfo{journal}{Phys. Rev. D} \bibinfo{volume}{74},
  \bibinfo{pages}{014020}.
\newblock \DOIprefix\doi{10.1103/PhysRevD.74.014020},
  \href{http://arxiv.org/abs/hep-ph/0606008}{{\tt arXiv:hep-ph/0606008}}.
%Type = Article
\bibitem[{Weinzierl(2008)}]{Weinzierl:2008iv}
\bibinfo{author}{Weinzierl, S.}, \bibinfo{year}{2008}.
\newblock \bibinfo{title}{{NNLO corrections to 3-jet observables in
  electron-positron annihilation}}.
\newblock \bibinfo{journal}{Phys. Rev. Lett.} \bibinfo{volume}{101},
  \bibinfo{pages}{162001}.
\newblock \DOIprefix\doi{10.1103/PhysRevLett.101.162001},
  \href{http://arxiv.org/abs/0807.3241}{{\tt arXiv:0807.3241}}.
%Type = Article
\bibitem[{Weinzierl(2009)}]{Weinzierl:2009nz}
\bibinfo{author}{Weinzierl, S.}, \bibinfo{year}{2009}.
\newblock \bibinfo{title}{{The infrared structure of $e^+ e^- \to 3$ jets at
  NNLO reloaded}}.
\newblock \bibinfo{journal}{JHEP} \bibinfo{volume}{07}, \bibinfo{pages}{009}.
\newblock \DOIprefix\doi{10.1088/1126-6708/2009/07/009},
  \href{http://arxiv.org/abs/0904.1145}{{\tt arXiv:0904.1145}}.
%Type = Article
\bibitem[{Zurbano~Fernandez et~al.(2020)}]{ZurbanoFernandez:2020cco}
\bibinfo{author}{Zurbano~Fernandez, I.}, et~al., \bibinfo{year}{2020}.
\newblock \bibinfo{title}{{High-Luminosity Large Hadron Collider (HL-LHC):
  Technical design report}} \bibinfo{volume}{10/2020}.
\newblock \DOIprefix\doi{10.23731/CYRM-2020-0010}.

\end{thebibliography}
